\documentclass[aps,prb,twocolumn,superscriptaddress,floatfix,showpacs,amsmath,amssymb,nofootinbib,longbibliography]{revtex4-2}

\usepackage{graphicx,color}
\usepackage{dcolumn}
\newcolumntype{.}{D{.}{.}{-1}}
\usepackage{bm}
\usepackage[utf8]{inputenc}
\usepackage{epstopdf}
\usepackage{amsmath,hyperref}
\usepackage{soul}
\usepackage{german}
\usepackage{nicefrac}
\captionsenglish
\usepackage{siunitx}             
\usepackage{xcolor}
\usepackage[normalem]{ulem}
\usepackage[capitalize]{cleveref}

\newcommand{\cpp}{$c_{\rm p}$}
\newcommand{\cpx}{$c_{\rm p}$}

\newcommand{\alvol}{$\alpha_{\rm vol}$}
\newcommand{\ala}{$\alpha_{\rm a}$}

\newcommand{\alc}{$\alpha_{\rm c}$}
\newcommand{\dllvol}{$dL_{\rm vol}/L_{\rm vol}$}

\newcommand{\ali}{$\alpha_{i}$}

\newcommand{\dlli}{$dL_{i}/L_{i}$}
\newcommand{\cpfe}{$c_{p}^{\mathrm{Li_2FeSiO_4}}$}
\newcommand{\cpzn}{$c_{p}^{\mathrm{Li_2ZnSiO_4}}$}

\newcommand{\rev}[1]{\textcolor{black}{#1}}

\newcommand{\cpm}{$c_p^{\rm non-ph}$}

\newcommand{\cpphon}{$c_p^{\rm ph}$}
\newcommand{\cpph}{$c_p^{\mathrm{ph}}$}
\newcommand{\cpo}{$c_p^{\mathrm{o}}$}

\newcommand{\cpnon}{$c_p^{\mathrm{non-ph}}$}
\newcommand{\cpmag}{$c_p^{\rm mag}$}

\newcommand{\cpsch}{$c_p^{\rm Sch}$}

\newcommand{\gci}{\gamma_c^{\mathrm{i}}}
\newcommand{\geffc}{\gamma_c^{\mathrm{eff}}}
\newcommand{\gcph}{\gamma_c^{\mathrm{ph}}}

\newcommand{\gco}{\gamma_c^{\mathrm{o}}}

\newcommand{\tn}{$T_{\rm N}$}

\newcommand{\lfs}{Li\(_2\)FeSiO\(_4\)}

\newcommand{\lzs}{Li\(_2\)ZnSiO\(_4\)}

\begin{document}

\title{Anisotropic magneto-elastic coupling in the quasi-2D Ising magnet Li$_2$FeSiO$_4$ and thermodynamic signature of low-energy orbital excitations}


\author{W.~Hergett}
\affiliation{Kirchhoff Institute of Physics, Heidelberg University, INF 227, D-69120 Heidelberg, Germany}

\author{K.~Ackermann}
\affiliation{Institute for Theoretical Physics, Heidelberg University, Philosophenweg 19, D-69120 Heidelberg, Germany}

\author{S.~Spachmann}
\affiliation{Kirchhoff Institute of Physics, Heidelberg University, INF 227, D-69120 Heidelberg, Germany}

\author{M.~Jonak}
\affiliation{Kirchhoff Institute of Physics, Heidelberg University, INF 227, D-69120 Heidelberg, Germany}

\author{M.~Abdel-Hafiez}
\affiliation{Department of Applied Physics and Astronomy, University of Sharjah, P. O. Box 27272 Sharjah, United Arab Emirates}
\affiliation{Physics Department, Faculty of Science, Fayoum University, Fayoum 63514, Egypt}

\author{Hend~Shahed}
\affiliation{Institute for Quantum Materials and Technologies, Karlsruhe Institute of Technology, Kaiserstr. 12, 76131 Karlsruhe, Germany}

\author{Michael~Merz}
\affiliation{Institute for Quantum Materials and Technologies, Karlsruhe Institute of Technology, Kaiserstr. 12, 76131 Karlsruhe, Germany}
\affiliation{Karlsruhe Nano Micro Facility (KNMFi), Karlsruhe Institute of Technology, Kaiserstr. 12, 76131 Karlsruhe, Germany}

\author{M.~W.~Haverkort}
\affiliation{Institute for Theoretical Physics, Heidelberg University, Philosophenweg 19, D-69120 Heidelberg, Germany}

\author{R.~Klingeler}
\email[Email:~]{klingeler@kip.uni-heidelberg.de}
\affiliation{Kirchhoff Institute of Physics, Heidelberg University, INF 227, D-69120 Heidelberg, Germany}


\date{\today}

\begin{abstract}
The relevance of spin, lattice and orbital degrees of freedom in the quasi-2D Ising-like magnet Li$_2$FeSiO$_4$ is demonstrated by thermal expansion and specific heat measurements supported by numerical studies. By means of thermodynamic and Gr\"{u}neisen analysis, we separate the orbital signatures and quantify the energy gap and its pressure dependence from the experimental data as $\Delta\sim 300 - 350$~K and $\partial \ln \Delta/\partial p_{\rm c} \approx -4.6$~\%/GPa). Both values excellently agree with the theoretical DFT predictions. The thermal expansion anomalies associated with the onset of long-range magnetic order imply significant magneto-elastic coupling and are used to determine the anisotropic uniaxial strain dependencies of $T_{\rm N}$. They are dominated by the in-plane (i.e., $c$ axis) response $(\partial \ln{ T_{\rm N}} )/\partial p_{\rm c} \simeq 15$~\%/GPa while the (negative) $a$ axis and (positive) out-of-plane ($b$ axis) pressure dependencies are much smaller. Short-range magnetic order extends well above $T_{\rm N}$ but is not driven by the same energy scale as $T_{\rm N}$ as evidenced by the observation of non-monotonic in-plane distortion, sign-change of the Gr\"{u}neisen parameter, and failure of Gr\"{u}neisen scaling.
\end{abstract}

\maketitle

\section{Introduction}

Transition‑metal compounds display a wealth of unexpected and spectacular phenomena -- including complex electronic phases and exotic quantum states -- originating from their intertwined charge, spin, lattice, and orbital degrees of freedom. Although charge and spin have been initially in the research focus, it has become increasingly evident that orbital physics is equally essential for deciphering and tuning the properties of these materials.~\cite{Tokura2000,Streltsov2017,Khomskii2022,Geck2004,Carrascolvarez2025} In transition metal compounds, the partially filled $d$ orbitals are split by the local crystal field, electron correlations, or spin-orbit coupling so that their degeneracy can be fully or partly lifted and orbital degrees of freedom become a relevant local electronic property. Associated orbital ordering phenomena are strongly linked to the crystal lattice, e.g., via Jahn–Teller distortions, leading to cooperative structural changes that stabilize particular orbital configurations.~\cite{Gehring1975} The orbital occupancy and arrangement also govern magnetic superexchange interactions such as Kugel-Khomskii coupling or the Goodenough–Kanamori-Anderson exchange
~\cite{Anderson1950,Goodenough1955,Kanamori1959,Kugel1973,Kugel1982} which further corroborates the orbital's role as a key ingredient of the complex physics of emergent phenomena in transition metal compounds.

Typically, orbital excitations in transition metal materials are studied via optical methods~\cite{Rueckamp2005,Ulrich2006, Benckiser2008} or resonant inelastic X-ray scattering (RIXS)~\cite{Murakami1998,Geck2005,Veenendaal2006,Schlappa2012}. Orbital degrees of freedom are also detected by thermodynamic properties, e.g., by entropy or lattice changes associated with short- or long-range orbital ordering phenomena or orbital rearrangements and redistribution.~\cite{Fritsch2002,Fritsch,Kriener2005,Klingeler2005,Sage2007,Stockert2012,Baek2014,Kumar2020,Ohlendorf2021,Qin2025,Hong2025} Here, we report thermal expansion and specific heat studies on the quasi-2D Ising-like magnet Li$_2$FeSiO$_4$. We find significant magneto-elastic coupling and quantify the uniaxial pressure dependencies. In addition to short-range spin-ordering evidenced by anisotropic lattice changes, our data reveal entropy changes associated with numerically predicted low-lying orbital excitations in \lfs\ which  are further confirmed by the temperature dependence of the Gr\"{u}neisen parameter. The numerical orbital configuration is corroborated by excellent agreement of the experimental and theoretically predicted uniaxial pressure dependence of the lowest orbital energy gap.
In summary, our data allow us to disentangle long- and short-range magnetic order, the lattice degrees of freedom, and the low-energy orbital excitations.



\section{Experimental}

The single crystals studied in this work have been grown by the high-pressure optical floating-zone method as described in detail in Refs.~\onlinecite{Hergett2019a,Hergett2019b,Hergett2021}. They crystallize in the $Pmnb$ phase and were further investigated in Refs.~\cite{Hergett2025,Hergett2026}. \rev{Composition, stoichiometry, and defect chemistry of the crystals have been characterized in detail in these previous studies. Single-crystal XRD, EDX, and high-resolution powder diffraction of ground crystals confirm the nominal stoichiometry and indicate only small ($\simeq$~2~\%) intrinsic Li–Fe antisite disorder~\cite{Hergett2019b}. 
}
The crystals were oriented by Laue diffraction and
cut perpendicular to the main crystallographic axes into cuboids with approximate dimensions \(1.1 \times 1.1 \times 1.0~\text{mm}^3\). Single-crystal XRD measurements were performed on a high-flux, high-resolution, rotating anode Rigaku Synergy-DW (Mo/Ag) diffractometer using Mo $K_{\alpha}$ radiation ($\lambda$ = 0.7107 \AA). The specimen had a size of $\approx$ 80 $\times$ 80 $\times$ 60 $\mu m^3$. The data were collected upon cooling from 310 to 80~K in temperature steps of 5~K and a cooling rate of 2~K/min. Data collection, integration of the intensities, and data reduction were performed using the most recent version of the \textsc{CrysAlisPro} software package~\cite{CrysAlis}.

High-resolution dilatometry measurements were performed using a three-terminal capacitance dilatometer from Kuechler Innovative Measurement Technology~\cite{kuechler2012}, operated inside a variable-temperature insert in an Oxford magnet system~\cite{Werner2017}. By means of the dilatometer, the uniaxial thermal expansion $dL_i(T)/L_i$ along the crystallographic axes ($i=a,b,c$) is determined which allows us to obtain the respective response functions, i.e., the linear thermal expansion coefficients $\alpha_i = 1/L_i \times dL_i(T)/dT$ at temperatures from 2 to 250~K \rev{(for $i=a$: 2 to 100~K)}. The volume thermal expansion is calculated as \dllvol~=~$1/3\times \sum_i{dL_i/L_i}$. \rev{The data were corrected for the cell effect of the dilatometer as obtained from a copper reference measurement, leaving at most a smooth, anomaly-free residual background.}

Following the approach outlined in Ref.~\cite{Hergett2026}, fully relativistic four-component density functional theory (DFT) calculations were performed using the FPLO code~\cite{FPLO1999}. The local density approximation (LDA) functional in the Perdew-Wang 92 (PW92) parameterization~\cite{PerdewWang1992} was employed, assuming a paramagnetic state. The Brillouin zone was sampled with a $20\times10\times20$ $k$-point mesh.
Effective local Hamiltonians were constructed by downfolding the self-consistent band structure onto maximally projected atomic-like Wannier functions~\cite{FPLOWAN2003}. To extract the strain dependence of the local orbital energy gap $\Delta$, the baseline crystal structure from Ref.~\cite{Hergett2019b} was strained along the crystallographic $c$-axis, while the remaining axes were scaled uniformly to conserve the total unit-cell volume.

\section{Experimental results}

\subsection{Thermal Expansion}

\begin{figure}[tb]
    \centering
    \includegraphics[width=1.0\columnwidth, clip]{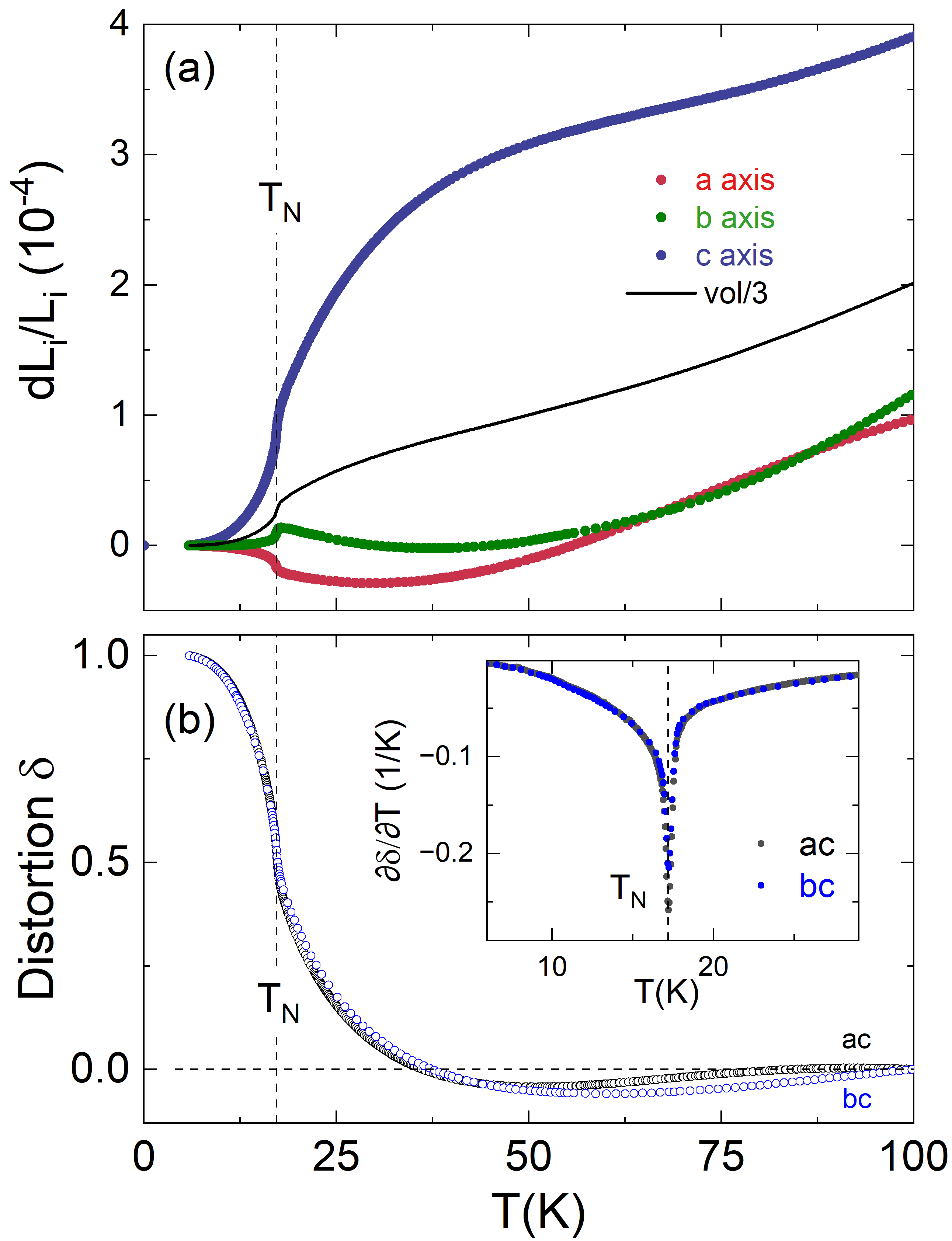}
    \caption{(a) Relative length changes, $dL_i(T)/L_i$ ($i = a, b, c$), of \lfs. The solid black line denotes the volume expansion \dllvol~=~$1/3\times \sum_i{dL_i/L_i}$; the vertical dashed line indicates \tn . (b) \rev{Distortion parameters $\delta_{ij} = (dL_\mathrm{i} - dL_\mathrm{j})/(dL_\mathrm{i} + dL_\mathrm{j})$ ($ij$ = $ac$ (black) and $bc$ (blue))} as a function of temperature. Inset: Temperature derivative $\partial \delta / \partial T$.
}
    \label{dll}
\end{figure}

\begin{figure}[tb]
    \centering
    \includegraphics[width=1.0\columnwidth, clip]{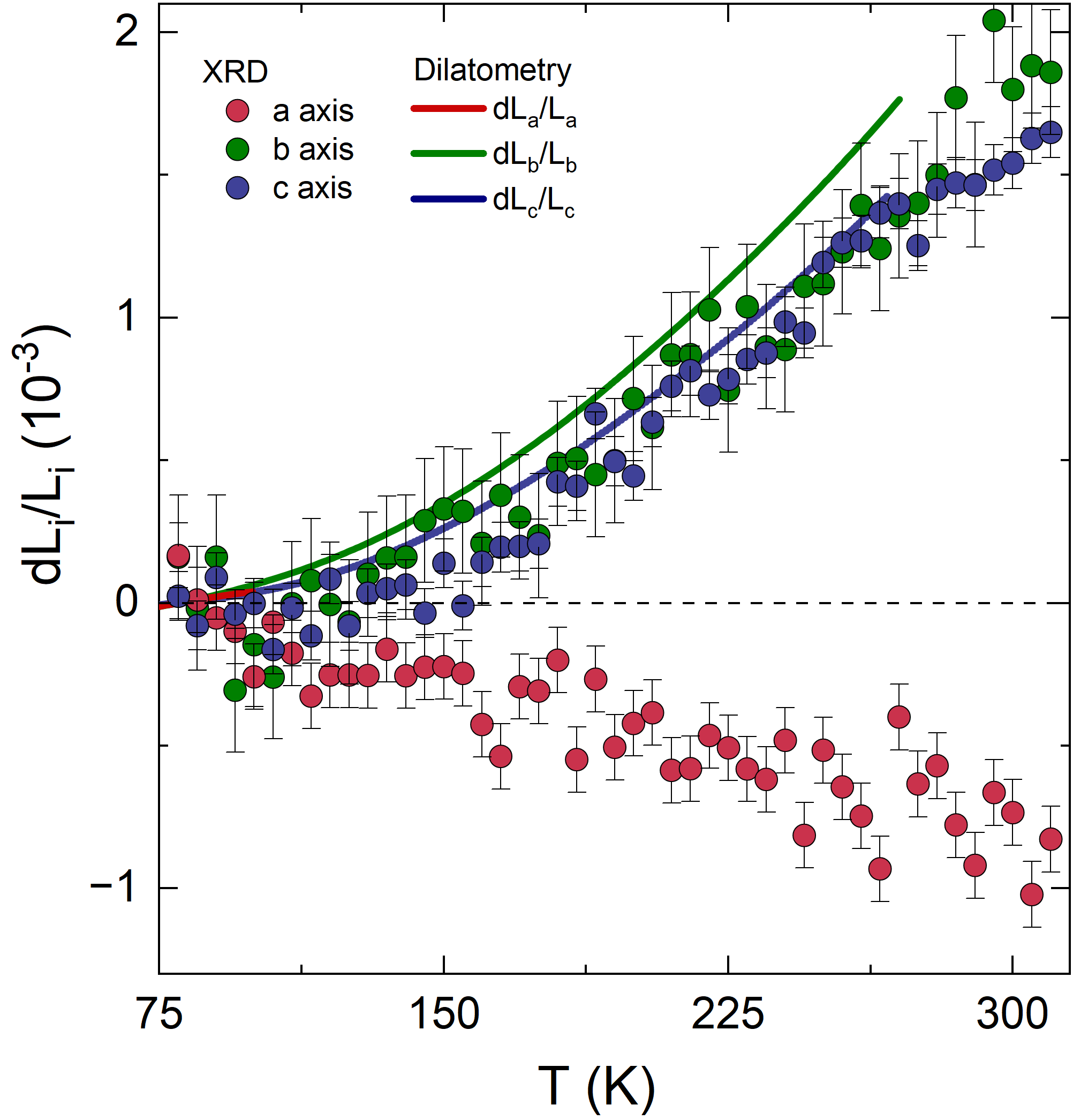}
    \caption{Relative length changes, $dL_i(T)/L_i$ ($i = a, b, c$), above 80 K as obtained from XRD and dilatometry. The dashed line indicates $\alpha_i(T\geq 80~{\rm K})=0$~K$^{-1}$. }
    \label{fig:xrd2}
\end{figure}

\begin{figure}[tb]
    \centering
    \includegraphics[width=1.0\columnwidth, clip]{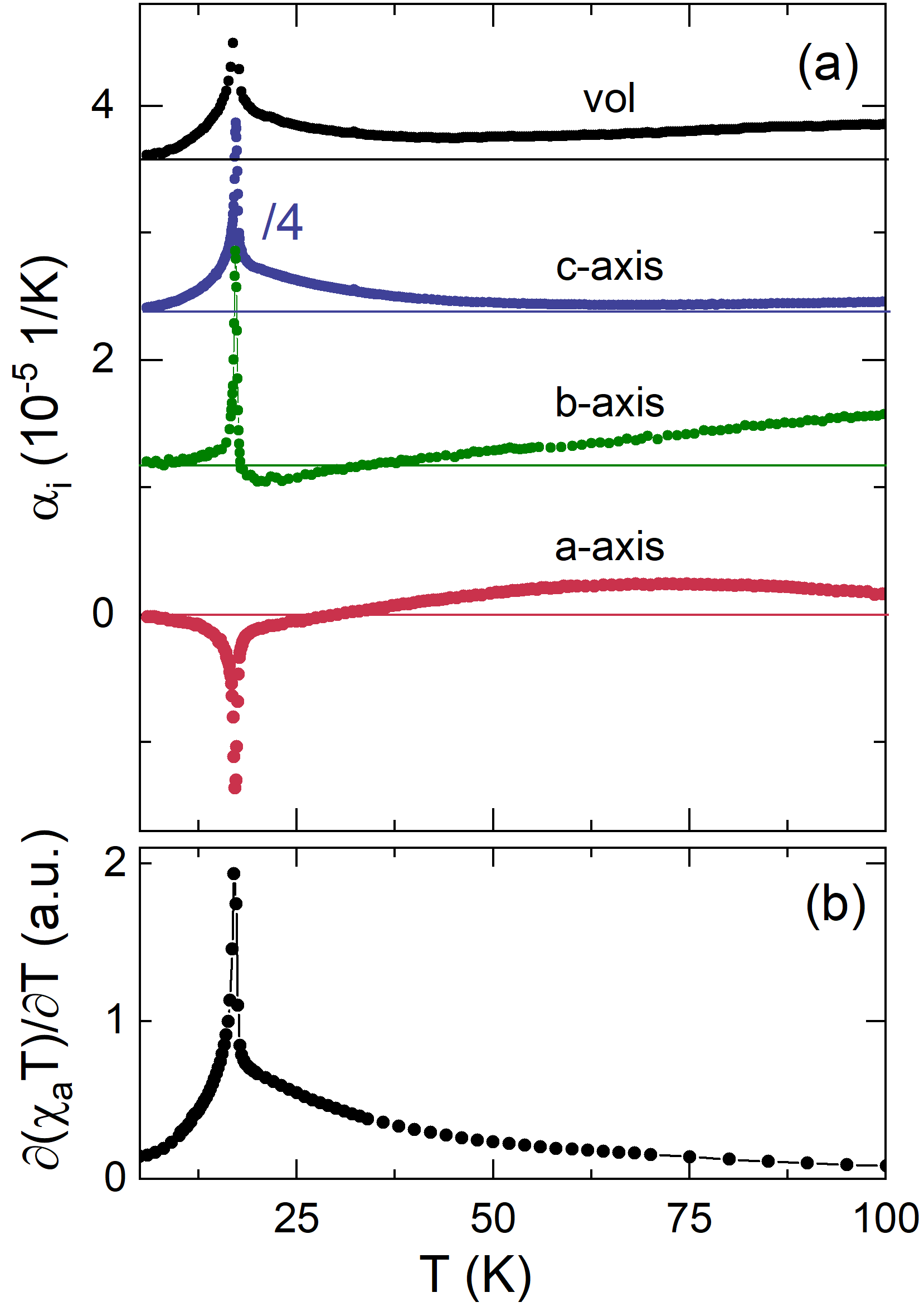}
    \caption{(a) Uniaxial thermal expansion coefficients \ali\ ($i$ = $a$, $b$, $c$, volume). \alc\ has been divided by 4. The data have been shifted vertically by $1.2\times 10^{-5}$/K for clarity; horizontal lines show $\alpha_{\rm i}=0$ for each data set. (b) Fisher's specific heat $\partial(\chi_{\rm a}T)/\partial T$ as derived from the static magnetic susceptibility $\chi_{\rm a}=M/B$ at $B=1$~T applied along the easy magnetic $a$ axis~\cite{Hergett2025}.
}
    \label{alpha_cp}
\end{figure}

The uniaxial and volume thermal expansion of \lfs\ presented in Fig.~\ref{dll} show a complex behavior and in particular provide information about the evolution of short- and long-range magnetic order in this material. Starting from around 100~K where all three crystallographic axes shrink upon cooling as shown in Figs.~\ref{dll} and in Fig.~\ref{fig:SM_dll_shift} in the Appendix, the following key features become apparent: (1) There is a wide temperature regime below $\sim 80$~K of strongly anisotropic thermal expansion as, e.g., indicated by the different sign of the slopes of the length changes \dlli\ along the different crystallographic axes $i=a,b,c$. Quantitatively, the overall length changes of the $c$ axis in the temperature regime $2 - 100$~K amount to $\Delta^{100\,{\rm K}}_{2\,{\rm K}} L_{\rm c}/L_{\rm c} \simeq 3.92 \times 10^{-4}$ which is four times larger than $\Delta^{100\,{\rm K}}_{2\,{\rm K}} L_{\rm a}/L_{\rm a} \simeq \Delta^{100\,{\rm K}}_{2\,{\rm K}} L_{\rm b}/L_{\rm b} \simeq 1 \times 10^{-4}$. (2) While the $c$ axis monotonically shrinks upon cooling, there are regions of negative thermal expansion in the $a$ and $b$ axis, respectively, yielding minima in both lattice parameters at 30 -- 40~K. (3) The observed strong anisotropy of the uniaxial thermal expansion extends to 300~K and above as evidenced by the temperature dependence of the lattice parameters (Fig.~\ref{fig:xrd1} in the Appendix) and is further illustrated by the relative length changes above 80 K as obtained from XRD and capacitance dilatometry Fig.~\ref{fig:xrd2}. 
The data imply that, upon heating, the $a$ axis shrinks (i.e., \ala~$<0$) \rev{above $\sim 115$~K} whereas the $b$ and $c$ axes expand in the temperature regime $\simeq 100$~K~$\leq T\leq 300$~K.~\footnote{\rev{$dL_{\rm a}/L_{\rm a}$ from capacity dilatometry could be obtained only for $T<100$~K. Within experimental uncertainty, it agrees to the XRD results. Note that $\alpha_{\rm a}$ decreases upon heating in this temperature regime (see Fig.~\ref{alpha_cp}).}} (4) There are clear kinks in the uniaxial and volume length changes at the long-range magnetic ordering temperature $T_{\rm N} = 17.2(5)$~K. These anomalies are visible in the length changes along all axes as well as in the volume thermal expansion \dllvol~=~$1/3\times \sum_i{dL_i/L_i}$. The data hence evidence significant magneto-elastic coupling in \lfs.

The anomalies at \tn\ are in particular visible in the thermal expansion coefficients \ali\ shown in Fig.~\ref{alpha_cp}a.~\footnote{$\alpha_{\rm vol}
=(\alpha_{\rm a}+\alpha_{\rm b}+\alpha_{\rm c})/3$.} We observe pronounced $\lambda$-shaped anomalies which imply the continuous nature of the phase transition. This is resembled by the anomaly in  Fisher's magnetic specific heat $c_{\rm p}^{\rm Fisher}=\partial(\chi_{\rm a}T)/\partial T$ derived from the static magnetic susceptibility along the easy magnetic axis. The contrasting signs and magnitudes of the thermal expansion anomalies along the different crystallographic axes, at \tn , reveal significant and anisotropic magneto-elastic coupling, manifested in a contraction of the $b$ and $c$ axes and an expansion of the $a$ axis upon the evolution of long-range magnetic order at \tn.

Notably, along the $b$ axis, the thermal expansion coefficient changes its sign to negative between \tn\ and approximately 40~K, highlighting the complex interplay between magnetism and lattice dynamics. \rev{For the quasi-2D material under study, we particularly note the in-plane anisotropy evolving upon cooling from 100~K where the thermal expansion along the $b$ and $c$ crystallographic axes develops pronounced additional anisotropy (see Figs.~\ref{dll}b and \ref{fig:SM_dll_shift}).} The most prominent effect is a pronounced shrinking of the $c$ axis below $\sim 50$~K with respect to the effects in the $ab$ plane.
Specifically, the $c$ axis shrinks upon cooling in the whole temperature regime under study, whereas the $a$ axis displays a broad minimum around $T\simeq 30$~K below which it expands towards \tn\ and beyond to the lowest temperatures.

The associated (in-plane) anisotropy in thermal expansion is illustrated by the distortion parameter $\delta_{ac} = (dL_\mathrm{a} - dL_\mathrm{c})/(dL_\mathrm{a} + dL_\mathrm{c})$ shown in Fig.~\ref{dll}b. Here, we consider only the length changes below 100~K so that $\delta_{ac}(100~{\rm K}) := 0$. The distortion is further illustrated in Fig.~\ref{fig:SM_dll_shift} in the Appendix where the relative length changes of the crystallographic axes below 100~K are shown. Upon cooling, $\delta$ slightly decreases down to about 50~K, i.e., the $c$ axis shrinks more than the $a$ axis. Below 50~K, this behavior reverses and $\delta_{ac}$ evolves finite positive values before saturating at low temperatures. \rev{Similar behavior is indicated by the analogously obtained parameter $\delta_{bc}$ (see Fig.~\ref{dll}b). At \tn , both $\delta_{ac}$ and $\delta_{bc}$ exhibit} a clear anomaly which appears $\lambda$-shaped in the temperature derivative $\partial\delta /\partial T$ (see the inset of Fig.~\ref{dll}b). It is straightforward to attribute the observed distortion to magneto-elastic coupling and spin degrees of freedom driving anisotropic thermal expansion in both the long-range and short-range spin-ordered regimes.

As demonstrated by the Ehrenfest relation $\partial T_{\rm N}/\partial p_i = T_{\rm N} V_{\mathrm{m}} \Delta\alpha_i/\Delta c_{\mathrm{p}}$ ($V_{\rm m}$ is the molar volume), for a continuous phase transition the jumps in $\alpha$ and \cpp\ at \tn\ imply the uniaxial pressure dependence of the transition temperature.~\cite{Klingeler.2006,Gegenwart.2016} For the pronounced $\lambda$-shaped anomalies observed here, the anomalies $\Delta\alpha_i$ and $\Delta c_{\rm p}$ however cannot be reliably determined from the experimental data. Since $\Delta c_{\rm p}$ is positive, the signs of the thermal expansion anomalies can still be used to determine the respective signs of the corresponding initial pressure dependencies. The negative anomaly observed along the $a$ axis hence shows that applying \textit{uniaxial} pressure $p||a$ will lower \tn, i.e., $\partial T_{\mathrm{N}}/\partial  p_{\mathrm{a}} < 0$. In contrast, the anomalies for the $b$ and $c$ axes indicate $\partial T_{\mathrm{N}}/\partial  p_{\mathrm{i}} > 0$ ($i = b, c$), with the effect being considerably larger along the $c$ axis. The anomaly in $\alpha_{\mathrm{vol}}$
is dominated by eminent uniaxial effects along the $c$ axis, hence indicating the positive \textit{hydrostatic} pressure dependence of \tn .

Typically, when phonon contributions are the dominant factor of the thermal expansion, $\alpha_{\rm i}$ are expected to increase monotonically in the temperature regime studied at hand. However, in \lfs, $\alpha_{\rm c}$ decreases up to around 60~K, where it reaches a plateau and converges with the values of $\alpha_{\rm a}$ and $\alpha_{\rm b}$, resulting in rather isotropic behavior at small values across all axes above $\sim 80$~K. This observation suggests that magneto-structural effects preceding the onset of long-range order at \tn\ extend to temperatures at least up to about 80~K.

\subsection{Low-energy orbital excitations}
\label{orb_ex}

\begin{figure}[tb]
    \centering
    \includegraphics[width=\columnwidth]{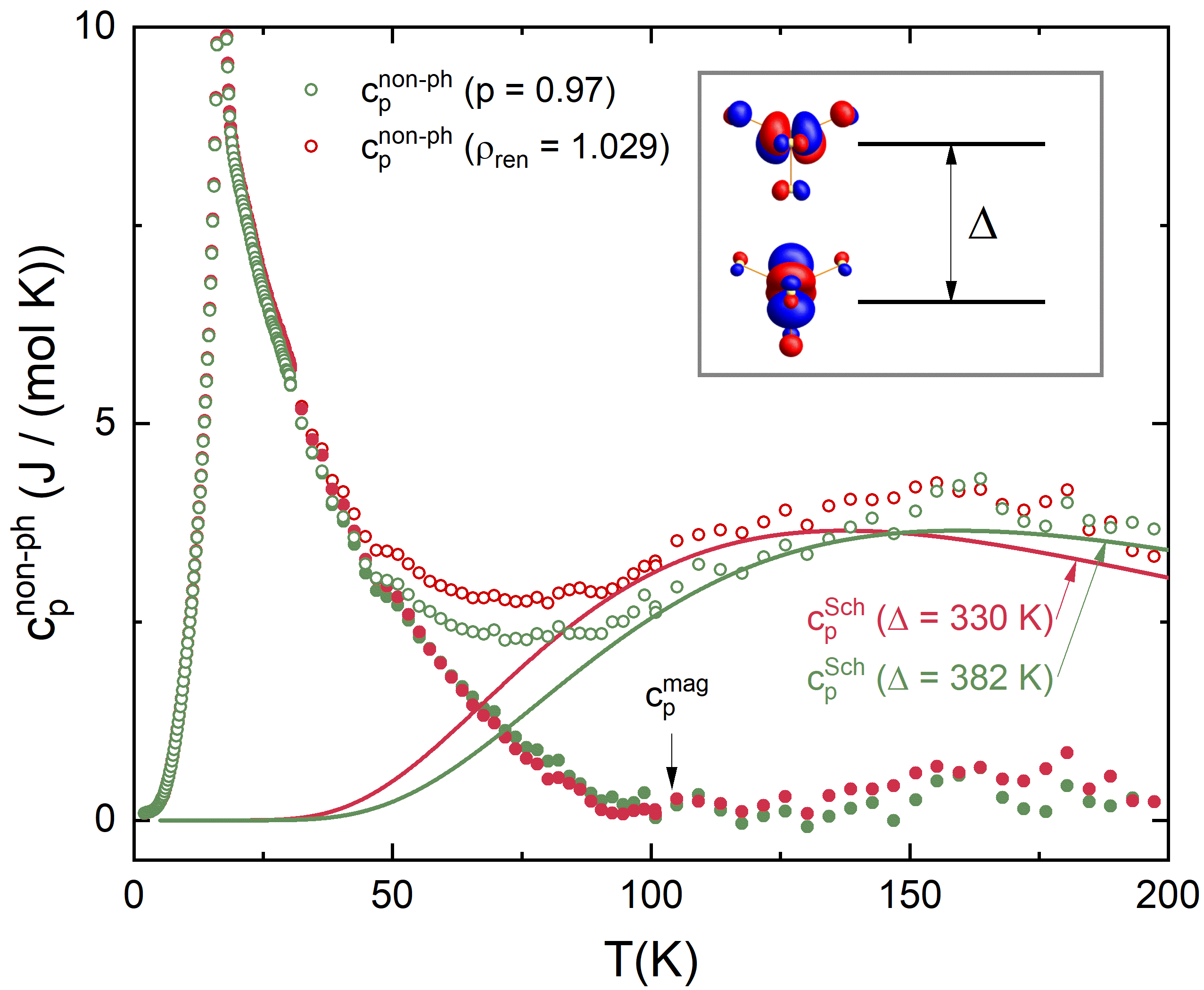}
    \caption{Separation of the non-phononic contributions \cpnon\ (i.e., the comprised magnetic and orbital specific heat) to the specific heat (open circles). Red data: \cpnon\ obtained by subtracting \cpzn\ from the measured specific heat using Debye scaling with $\rho_{\mathrm{D}} \approx 1.029$ (see the text). The red line is a fit to the high-temperature regime of the thus obtained \cpnon\ by means of the Schottky model, resulting in the energy gap $\Delta = 330$~K. Green data: The line represents \cpsch\ calculated by Eq.~\ref{eq:schottky} using the DFT result $\Delta = 382$~K~\cite{Hergett2026}. This \cpsch\ was used to determine the scaling factor $p=0.97$ so that \cpnon\ = \cpsch\ + \cpmag~(i.e., \cpnon~$\approx$~\cpsch\ in the regime where the spin correlations fade out).  For both approaches, the magnetic specific heat \cpmag~=~\cpnon~-~\cpsch\ is shown by the filled data points.
    Inset: Isosurfaces of the two lowest crystal-field/SOC-split Fe-centered $3d$-like Wannier states, chosen to enclose 90\% of the electronic density.}
    \label{cpfisher}
\end{figure}

The presence of short-range magnetic order well above \tn\ evidenced by our thermal-expansion data is also reflected by  Fisher's specific heat (see Fig.~\ref{alpha_cp}b) which shows magnetic entropy changes up to about 80~K. \rev{Magnetic entropy changes are also clearly visible in the specific heat of \lfs .~\cite{Hergett2025} In the following, we extend the phenomenological approach from Ref.~\cite{Hergett2025} to separate the phonon and non-phonon entropy changes by quantitative Debye scaling of the specific heat of \lzs\ which is the non-magnetic analog of the title compound (cf., e.g., Ref.~\cite{Wang2012}; for details see the Appendix). Using Debye approximation, the phonon heat capacity (\cpphon ) is obtained by Debye-scaling of \cpzn\ (temperature-scaling factor $\rho_{\mathrm{D}} = 1.029$) which subtraction from \cpfe\ yields the non-phonon heat capacity (\cpnon ). The latter hence comprises all entropy changes associated with the Fe-subsystem  (see Fig.~\ref{cpfisher}). Notably, in addition to pronounced short- and long-range magnetic order entropy changes, the data in Fig.~\ref{cpfisher} imply finite \cpnon\ at high temperatures which even increases upon heating above $\sim75$~K. While such a behavior cannot be expected for the spin degrees of freedom, it is tempting to attribute this feature to the orbital electronic structure.}







It has been found by density functional calculations~\cite{Hergett2026} that the electronic structure in \lfs\ is characterized by energetically isolated and rather flat bands near the Fermi level of mainly iron 3$d$ orbital character. Notably, the orbital degeneracy is found to be completely lifted by the combined effect of the crystal-field and spin-orbit coupling.
The full relativistic calculation yields $\Delta \simeq 28 - 33$~meV ($\simeq 325 - 380$~K) for the energy gap between the two lowest orbital levels which are well separated ($>300$~meV) from the next orbital states~\cite{Hergett2026}. One hence must expect associated entropy changes peaking around 150~K which straightforwardly explains the excess specific heat obtained by employing the quantitative scaling procedure described above. The 'non-phononic' heat capacity hence comprises sizable contributions from both the spin and the orbital degrees of freedom. The associated 'orbital' heat capacity can be described in terms of a two-level Schottky model,

\begin{equation}
c_{\rm p}^{\mathrm{Sch}}(T)
=
R
\left(\frac{\Delta}{T}\right)^2
\frac{e^{-\Delta/T}}
{\left(1+e^{-\Delta/T}\right)^2}.
\label{eq:schottky}
\end{equation}

\rev{Note that the energy gap $\Delta$ is the only free parameter. Fitting the high-temperature regime of \cpnon\ by means of Eq.~\ref{eq:schottky} yields a good description of the data (see Fig.~\ref{cpfisher}) and results in $\Delta^{\rm (exp)}=330(10)$~K. While the presence of $\Delta$ can be unambiguously deduced from the data, the experimental uncertainties of specific heat studies on two different samples (e.g., error in the mass determination) render its precise determination challenging. In particular, the experimental data do not exclude the numerically predicted value $\Delta^{\rm(th)} = 382$~K (32.9~meV) from Ref.~\cite{Hergett2026} as illustrated in Fig.~\ref{cpfisher} by the green data points. Here, we have described the excess specific heat in terms of Eq.~\ref{eq:schottky} using $\Delta^{\rm(th)}$ while the phonon contribution \cpph = $p\,$\cpzn\ was then adjusted by using the scaling factor $p$ as a fit parameter. This yields $p=0.97$ (see Fig.~\ref{cpfisher}) which within error bars agrees well with the approximated Debye scaling factor ($\rho_{\mathrm{D}}=1.029$). Hence, both approaches yield a good description of the experimental data and provide parameters which are in the regime of the expected numerical and experimental uncertainties. Estimating all uncertainties, our specific heat data imply that $\Delta$ is within the regime 310 -- 400~K.}



\rev{Our estimate of the orbital entropy changes also allows us to identify the bare magnetic contribution to the specific heat. For both approaches presented above, i.e., fixing either the Debye scaling parameter or the numerically predicted gap value, }
we find that the residual \cpfe-\cpphon-\cpsch\ agrees very well with  Fisher's specific heat while smoothly fading for $T \gtrsim 100$~K \rev{(see Fig.~\ref{fig:appendix-fisher} in the Appendix). This excellent agreement further confirms our quantitative approach to separate the phonon, magnetic, and orbital entropy changes.}

\begin{figure}[htb]
    \centering
    \includegraphics[width=\columnwidth]{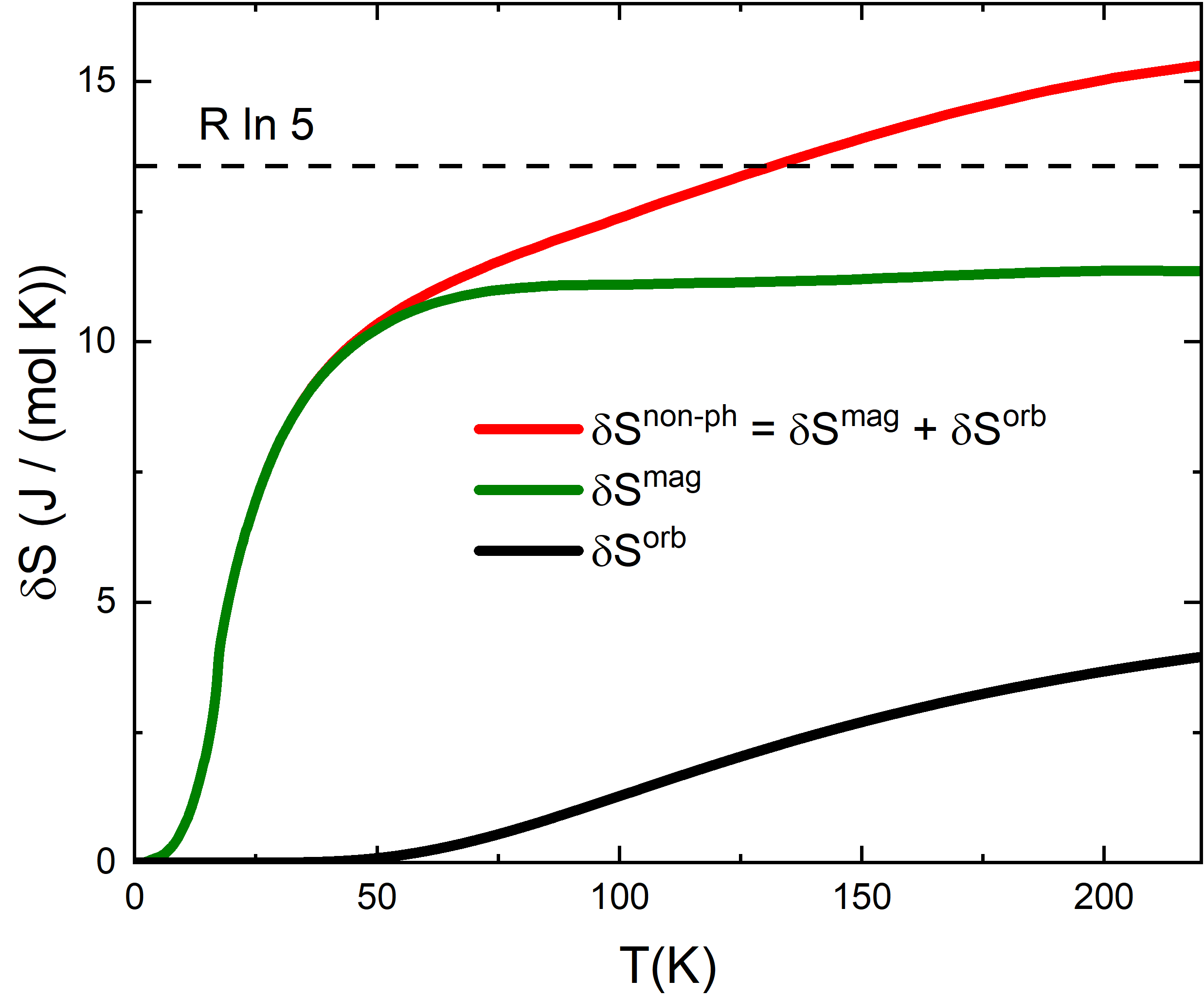}
    \caption{Temperature dependence of the entropy changes obtained by integrating the heat-capacity contributions shown in Fig.~\ref{cpfisher}. The total non-phononic entropy changes, $\delta S^{\mathrm{non\mbox{-}ph}}=\delta S^{\mathrm{mag}}+\delta S^{\mathrm{orb}}$, are shown in red. The green curve represents the magnetic entropy changes, $\delta S^{\mathrm{mag}}$, obtained by integrating \cpm-\cpsch, while the black curve shows the corresponding orbital entropy, $\delta S^{\mathrm{orb}}(T)=\int_{0}^{T}\!\left(c_p^{\mathrm{Sch}}(\tilde{T})/\tilde{T}\right)\,d\tilde{T}$. The dashed horizontal line indicates the reference value $R\ln 5$, expected for high-spin Fe$^{2+}$ ($S=2$).}
    \label{entropy}
\end{figure}


The temperature dependence of the calculated entropy changes shown in Fig.~\ref{entropy} provides further thermodynamic support for separating the non-phononic response into spin and orbital contributions. The overall non-phononic entropy changes $\delta S^{\mathrm{non-ph}}(T)=\int_{0}^{T}\!\left(c_p^{\mathrm{non-ph}}(\tilde{T})/\tilde{T}\right)\,d\tilde{T}$ clearly exceed the total spin entropy of $R\ln 5$ expected for high-spin Fe$^{2+}$ ($S=2$). This implies the presence of  additional degrees of freedom which is expected by the numerically predicted orbital configuration. The decomposition of $\delta S^{\mathrm{non\mbox{-}ph}}$ reveals two distinct temperature regimes. Below $\sim 50$~K the steep increase of the entropy changes suggests their magnetic nature. Calculating the magnetic entropy changes $\delta S^{\mathrm{mag}}$
yields the green curve which saturates at $\approx 11.4~\mathrm{J\,mol^{-1}\,K^{-1}}$. This is in the vicinity but slightly below the full spin-only value, thereby indicating that most of the spin entropy is released at 2~K~$\leq T \lesssim 4\times T_{\rm N}$. 
In contrast, the orbital contribution, $\delta S^{\mathrm{orb}}(T)$, is negligible at low temperatures and starts to increase above $\sim 60$~K, showing a broad and gradual recovery typical of a thermally activated Schottky-like excitation. At 220~K, $\delta S^{\mathrm{orb}}(T)$ reaches $\approx 4~\mathrm{J\,mol^{-1}\,K^{-1}}$, remaining below the limiting value of a two-level system $R\ln 2$. 
Overall, the analysis of the non-phononic entropy changes supports a picture in which the low-temperature thermodynamics are governed predominantly by spin degrees of freedom, whereas the additional entropy changes above $4\times T_{\rm N}$ are associated with the  orbital degree of freedom.

We finally note that the corrected background determination of the specific heat by using the $\rho_{\mathrm{D}}$ as described above neither affects the low-temperature analysis of the critical behavior nor the conclusions of dimensionality crossover drawn in Ref.~\cite{Hergett2026}. The analysis of \cpnon~ from Fig.~\ref{cpfisher} yields the critical exponents $\tilde{\alpha}^{\rm cp}=0.11(3)$ (around \tn ) and $\tilde{\alpha}^{\rm cp}=0.03(2)$ ($\gtrsim 22$~K), thereby confirming the crossover from 2D Ising-like to 3D Ising behavior below $\simeq 1.3\times T_{\rm N}$ (see Fig.~\ref{app:critical}).

\subsection{Gr\"{u}neisen scaling and uniaxial pressure effects\label{ch:gruen}}

Further information on the nature of short- and long-range ordering phenomena is obtained by comparing the entropy and the lattice changes. In general, separation and characterization of various degrees of freedom is feasible
through the Gr\"{u}neisen parameters $\gamma_i =$~\ali/\cpx . At the phase boundary of a continuous phase transition, the pressure dependencies of the associated long-range ordering phenomenon are obtained by comparing the anomalies at the phase transition via the Ehrenfest relation or -- in case of pronounced $\lambda$-character of the anomalies as present at hand -- by applying a method proposed by Pippard~\cite{pippard1957} and further developed by Souza \textit{et al.}~\cite{Souza2005}.

\subsubsection{Uniaxial and hydrostatic pressure dependence of \tn }

\begin{figure}[b]
    \centering
    \includegraphics[width=1.0\columnwidth, clip]{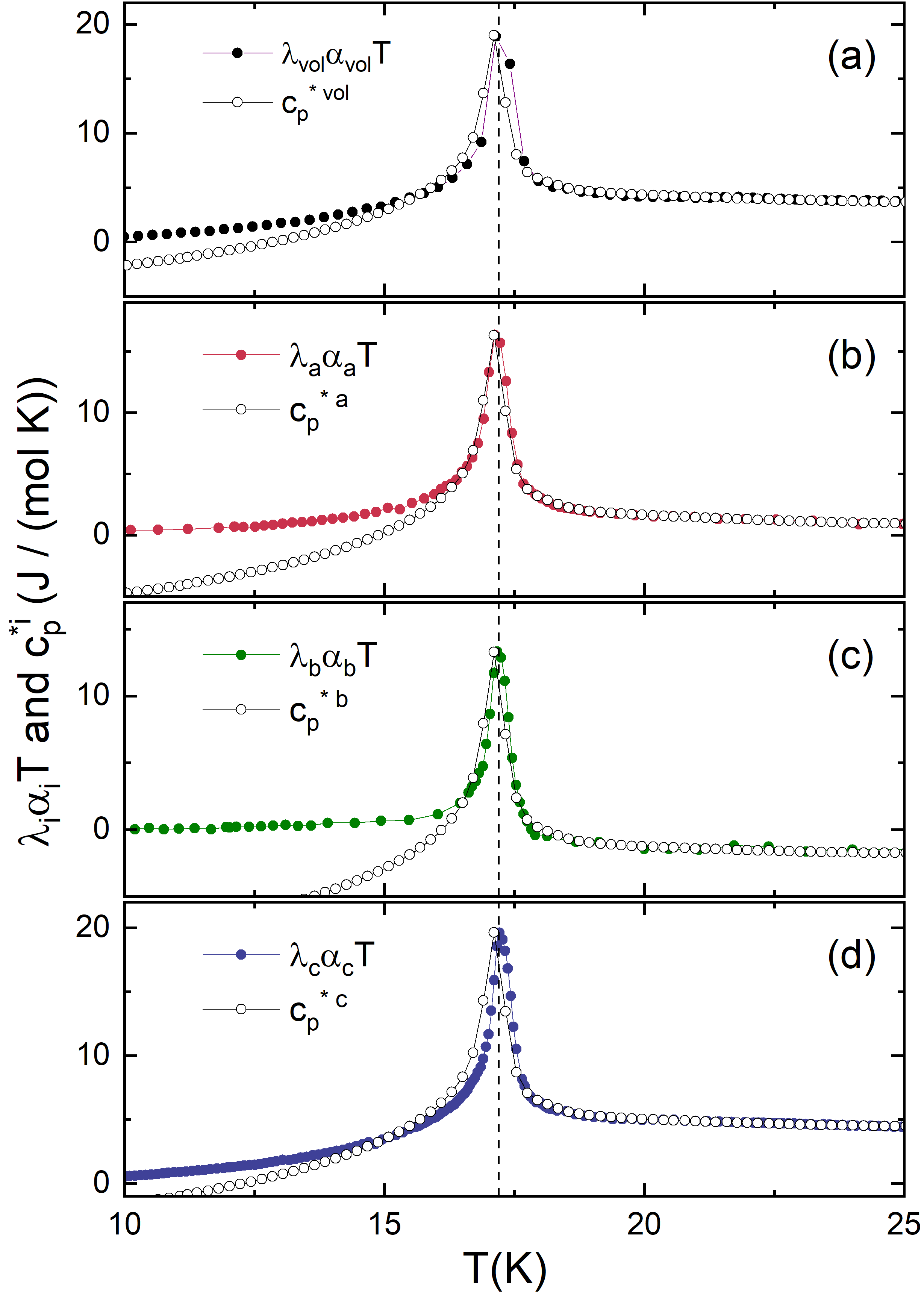}
    \caption{Comparison of $\lambda_i \alpha_i T$ and $c_p^{\ast i}$ in the vicinity of the phase transition for the volume and the three crystallographic axes ($i=a,b,c, vol$). Here, $\lambda_i$ is a scaling factor and $c_p^{\ast i}$ denotes the specific heat capacity after subtraction of the linear term $(A_i+B_iT)$ (see the text). The dashed vertical line marks \tn.}
    \label{lambda_alpha_cp}
\end{figure}

While the sign of the pressure dependencies at \tn\ can be straightforwardly deduced from the anomalies in \ali\ at \tn\ as discussed above, the pronounced $\lambda$-shape of the anomalies hinders the reliable determination of the jumps in \ali\ and \cpp , at \tn , and hence prevents the quantitative estimate of $\partial T_{\rm N}/\partial p_i$ by means of the Ehrenfest relation. Instead, the uniaxial pressure dependence of the ordering temperature can be evaluated using the approaches by Pippard~\cite{pippard1957} and by Souza \textit{et al.}~\cite{Souza2005}. \rev{Pippard argued that, close to a $\lambda$-transition, the entropy surface $S(T,p)$ exhibits a fold along the phase boundary and may be approximated by a cylindrical surface~\cite{pippard1957}. 
The rapidly varying part of $S(T,p)$ depends on temperature and pressure only through the distance to the (locally straight) transition line, while the entropy on the transition line varies smoothly with pressure. Pressure thus enters the rapidly varying part of the entropy only by shifting \tn($p_i$). In the language of Gr\"{u}neisen scaling this corresponds to a single energy scale with a single pressure dependence~\cite{Gegenwart.2016}. As a consequence, the anomalous contributions to $c_{\rm p}$ and to $\alpha_i T$ are proportional to each other, while all smoothly varying contributions are absorbed in a background term so that the uniaxial pressure effects are given by}

\begin{equation}
\frac{\partial T_{\rm N}}{\partial p_i} = \frac{V_m \alpha_i T}{c_{\rm p}^{\ast}}.
\label{eq:souza}
\end{equation}

Here, 
\rev{$c_{\rm p}^{\ast} = c_{\rm p} - (A_i + B_iT)$, with the
background linearized over the narrow fitting window ($A_i$ is a constant and $B_i$ is the linear coefficient).} Application of Eq.~\eqref{eq:souza} requires that the $c_{\rm p}^{\ast}$ and $\lambda \alpha T$ curves match in the vicinity of \tn . The scaling factor \rev{$\lambda_i$} as well as the parameters \rev{$A_i$} and \rev{$B_i$} are therefore adjusted to achieve the best possible agreement between the two curves (see Fig.~\ref{lambda_alpha_cp}).
The resulting parameters are summarized in Table~\ref{tab:souza_parameters}. The obtained pressure dependencies confirm that the increase of \tn\ upon application of hydrostatic pressure $p_h$ is driven by the large effects of in-plane uniaxial pressure $p||c$ axis while in-plane pressure $p||a$ axis yields a much smaller decrease of \tn . Out-of-plane pressure ($p||b$ axis) moderately stabilizes long-range magnetic order. Quantitatively, 
the pressure coefficients amount to $(\partial \ln T_{\rm N})/\partial p_i \simeq$ 14.5(5), -4.2(1), 6.3(2), and  15.2(7)~\%/GPa ($i={\rm vol},a,b,c$).\footnote{$\frac{dT_{\rm N}}{dp_{\rm h}}= \frac{\partial T_{\rm N}}{\partial p_a}
+\frac{\partial T_{\rm N}}{\partial p_b}
+\frac{\partial T_{\rm N}}{\partial p_c}=\frac{3V_m}{\lambda_{\rm vol}}.$}

\rev{Evaluated at $T \approx$~\tn, Eq.~\eqref{eq:souza} coincides with the Gr\"{u}neisen relation (Eq.~\eqref{eq:single}) for $\varepsilon = k_{\rm B} T_{\rm N}$. The Souza-Pippard analysis is thus the critical-regime realization of Gr\"{u}neisen scaling. This equivalence also delineates the limits of the approach. If several degrees of freedom with different pressure dependencies contribute, the effective Gr\"{u}neisen parameter becomes a weighted average of the individual contributions (Eq.~\eqref{eq:gamma_eff}). The anomalous contributions to $c_{\rm p}$ and $\alpha_i T$ cease to be proportional, and no single scaling factor can superimpose them. The quality of the scaling at \tn\ (Fig.~\ref{lambda_alpha_cp}) hence evidences a single dominant energy scale at the transition, whereas the failure of the analogous low-temperature Gr\"{u}neisen scaling for the $a$ and $b$ axes at $T > $~\tn\ (Sec.~\ref{lro_sro}) signals the presence of several competing energy scales in the short-range ordered regime.  }

\begin{table}[htb]
\centering
\caption{Parameters obtained from the scaling analysis described in the text, together with the calculated uniaxial and hydrostatic ('vol') pressure dependencies of \tn .}
\label{tab:souza_parameters}
\setlength{\tabcolsep}{5pt}
\renewcommand{\arraystretch}{1.05}
\small
\begin{tabular}{@{}lcccc@{}}
\hline\hline
 & $\lambda_i$ & $A_i$ & $B_i$ & $\partial T_{\rm N}/\partial p_i$ \\
 & $(\mathrm{kJ}/(\mathrm{mol\,K}))$ & $(\mathrm{J}/(\mathrm{mol\,K}))$ & $(\mathrm{J}/(\mathrm{mol\,K^2}))$ & $(\mathrm{K}/\mathrm{GPa})$ \\
\hline
$a$  & -70(1)  & 6.2  & 0.06  & -0.72(1)  \\
$b$  & 46.8(7)    & 10.0 & 0.015 & 1.08(2)   \\
$c$  & 19.4(7)    & 3.4  & 0.03  & 2.61(9)  \\
vol    & 61(1)   & 3.7  & 0.05  & 2.49(4)   \\
\hline\hline
\end{tabular}
\end{table}

\subsubsection{Magnetic long- and short-range order}
\label{lro_sro}

In general, and beyond the thermodynamic phase boundaries, the pressure dependence of the relevant degrees of freedom can be accessed via the Gr\"{u}neisen ratio, $\gamma_i =$~\ali/\cpx , which enables the identification and characterisation of key energy scales. If a system is determined by a single energy scale $\varepsilon$ (or after separating the different  contributions to the specific and the thermal expansion coefficients), the Gr\"{u}neisen parameter $\gamma^{\mathrm{\varepsilon}}$ provides the associated (uniaxial or hydrostatic) pressure dependencies, i.e.,

\begin{equation}
\gamma^{\mathrm{\varepsilon}}_i = \frac{1}{V_m}\left. \frac{\partial\ln{\varepsilon}}{\partial p_i}\right|_T.
\label{eq:single}
\end{equation}

In complex systems as presented here, the various contributions to heat capacity and thermal expansion may be difficult to discriminate. In this case, the data provide a Gr\"{u}neisen parameter which is the sum of the Gr\"{u}neisen parameters from the various contributions times their relative fraction to the total heat capacity~\cite{deVisser1990,BarronWhite1999,Gegenwart.2016}.

\begin{equation}
\gamma^{\mathrm{eff}}
=
\frac{\sum_j \gamma_j\, c_j}{\sum_j c_j}. \label{eq:gamma_eff}
\end{equation}

\begin{figure}[tb]
    \centering
    \includegraphics[width=1\columnwidth, clip]{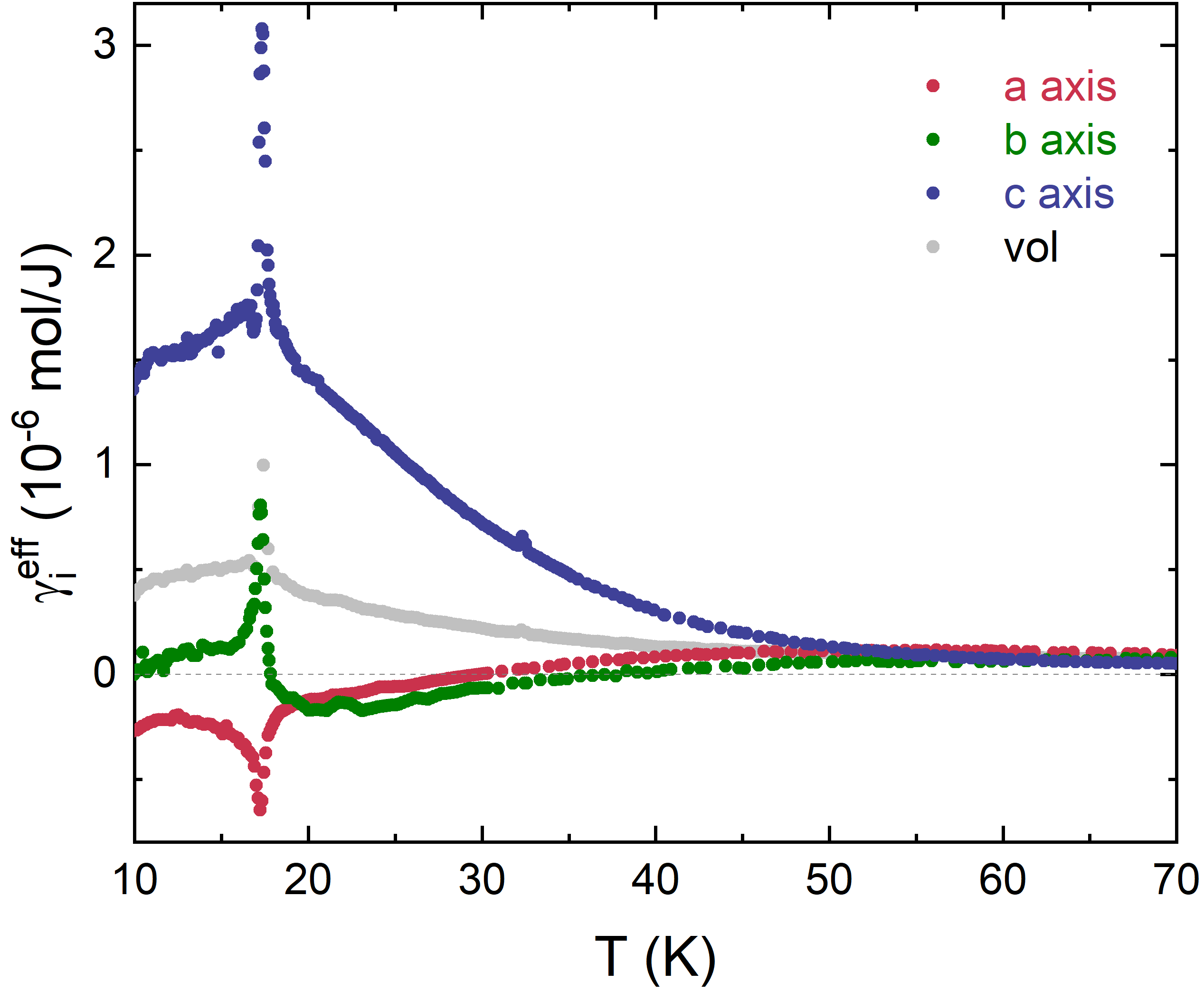}
    \caption{Gr\"{u}neisen parameters $\gamma^{\rm eff}_{\rm i}=\alpha_{\rm i}/c_{\rm p}$ for the three crystallographic axes $i=a,b,c$ and the volume. $\gamma^{\rm eff}_{\rm i}$ is dominated by the lattice and orbital degrees of freedom at high temperatures and by magnetic degrees of freedom at low temperatures.
    }
    \label{fig:totalgruen}
\end{figure}

Firstly, we consider the low temperature regime where phonon and orbital effects are supposed to be comparably small. Fig.~\ref{fig:totalgruen} presents the temperature dependence of the effective uniaxial Gr\"{u}neisen parameters $\gamma^{\rm eff}_i(T)$, where the experimentally measured total thermal expansion and specific heat are exploited. Above $\sim 60$~K, $\gamma^{\rm eff}_i(T)$ assumes comparably very small and rather temperature-independent values. The corresponding Gr\"{u}neisen ratio at 100~K is $\gamma^{\rm eff}_{a,b,c} \simeq 6(3) \times 10^{-8} \, \text{mol/J}$ which we mainly attribute to phonons and orbital excitations and will be discussed below.

Below 60~K, distinct behaviors emerge along the three crystallographic axes, reflecting the presence of magnetic degrees of freedom and anisotropic magnetic coupling associated with opposite uniaxial pressure dependencies along the different crystallographic directions. Specifically, $\gamma^{\rm eff}_c(T)$ increases upon cooling, showing the strongest variation and a dominant anomaly at \tn. In general, for analyzing the spin degrees of freedom, the \textit{magnetic} Gr\"{u}neisen parameter needs to be considered, i.e., the ratio of the purely magnetic contributions to the thermal expansion coefficient and the specific heat. However, the phonon background contribution to the thermal expansion coefficients is negligible in the relevant temperature regime below 30~K as indicated by Fig.~\ref{fig:TE-background} in the Appendix. The 'anomalous' contributions (i.e., those associated with the phase transition at \tn\ and with  precursing short-range order) of the thermal expansion coefficients, hence, are by far dominated by non-phononic degrees of freedom so that correcting $\alpha$ for an estimated background does not improve the following analysis. This can also be directly concluded from the large regime of decreasing \alvol\ and the anisotropy in \ali\ which both are not expected for a non-magnetic system (Fig.~\ref{alpha_cp}).\footnote{Note that orbital excitations are frozen out in this temperature regime, too.} In contrast, there is a sizable phonon contribution to the specific heat~\cite{Hergett2025} even at low temperatures. We hence approximate the magnetic Gr\"{u}neisen ratio (which in this temperature regime agrees with the non-phononic one) by $\gamma^{\rm mag}_i = \alpha_{i}^{\rm mag}/$\cpm~$\simeq$~\ali/\cpm . The completely different significance of the phonon background in the specific heat and the thermal expansion coefficient can be straightforwardly explained by the strong differences in the pressure dependencies of the relevant (magnetic and phononic) energy scales \rev{and hence of the Gr\"{u}neisen parameters $\gamma^{\rm mag}_c \simeq 1.8\times 10^{-6}$~mol/J and $\gcph \simeq 1.5 \times 10^{-7}\,\mathrm{mol/J}$ illustrated in Fig.~\ref{schottky_gamma}} (see also Fig.~\ref{fig:TE-background} in the Appendix). The observation that the total Gr\"{u}neisen parameters $\gamma^{\rm eff}_i =$~\ali /\cpp\ presented in Fig.~\ref{fig:totalgruen} are very small and rather temperature-independent around 70~K, where phonon contributions play a much larger role than around \tn , strongly supports our conclusions.

\begin{figure}[bt]
    \centering
    \includegraphics[width=1.0\columnwidth, clip]{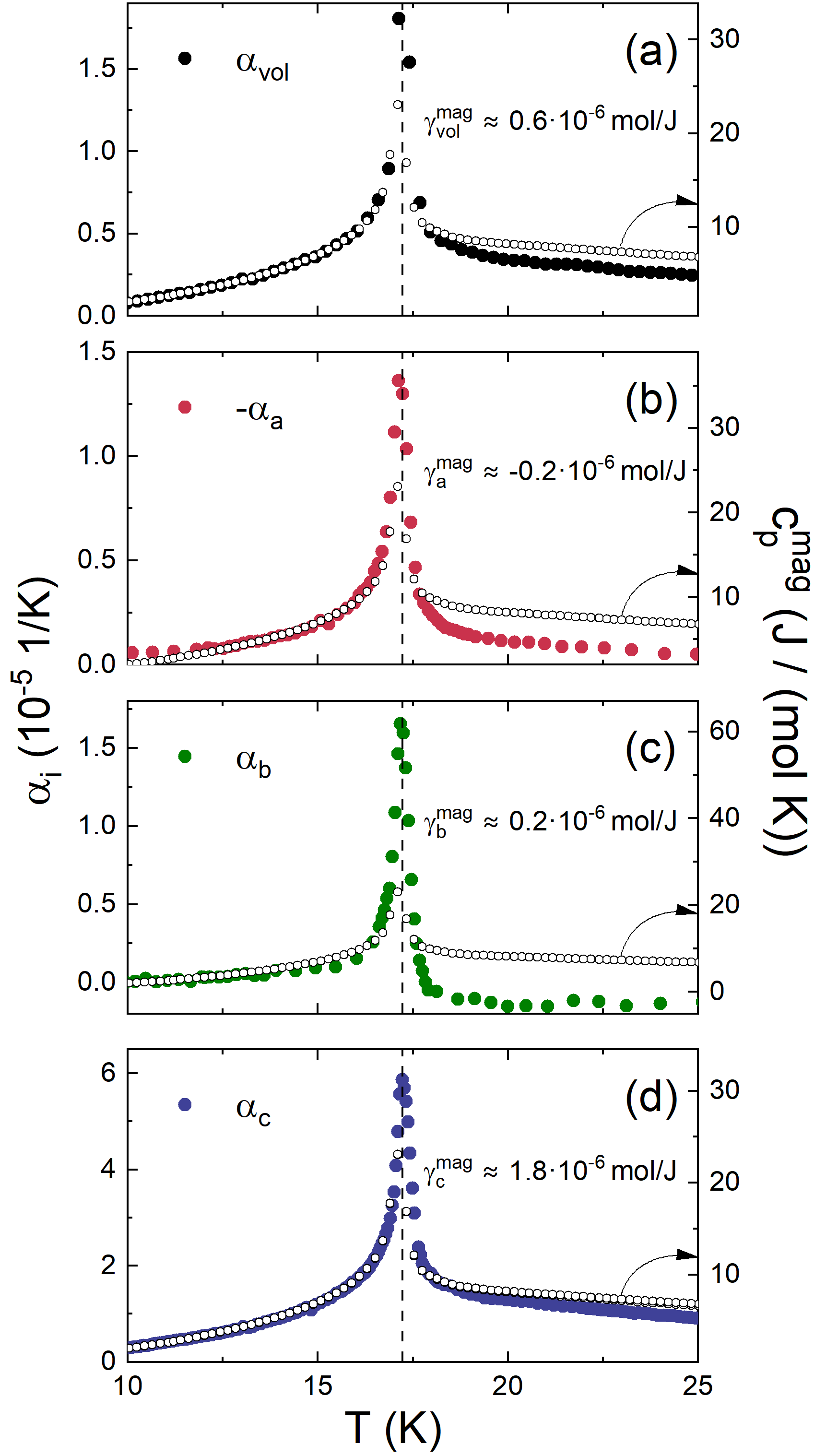}
    \caption{Thermal expansion coefficients (left ordinate; note the sign in (b)) and magnetic specific heat (right ordinate) scaled to visualize potential Gr\"{u}neisen scaling. $\gamma^{\rm mag}_i \simeq$~\ali/\cpm\ are the magnetic Gr\"{u}neisen parameters reflecting the scaling of the left and right ordinates.
    }
    \label{fig:gruen2}
\end{figure}

At low-temperatures below \tn , our data demonstrate a nearly temperature-independent Gr\"{u}neisen ratio across all three crystallographic directions.  In order to exploit the proportionality of \ali\ and \cpm\ in the long-range ordered phase, we have hence plotted both quantities with appropriate scaling so that the overlap at $T < T_{\rm N}$ is maximized. As shown in Fig.~\ref{fig:gruen2}, there is an excellent agreement of both quantities in the magnetically ordered phase which reasonably well extends, for the volume Gr\"{u}neisen ratio as well as for the $c$ axis, also to $T>T_{\rm N}$. The plot in particular shows that the magnetic Gr\"{u}neisen parameters $\gamma^{\rm mag}_i \simeq$~\ali/\cpm\ are constant in the magnetically ordered phase. Following Eq.~\eqref{eq:single}, this shows the presence of one dominant energy scale whose pressure dependencies can be read off from the Gr\"{u}neisen ratios.

Notably, the data in Fig.~\ref{fig:gruen2} show nearly perfect scaling for the $c$ axis, i.e., \alc~$\propto$~\cpmag , even above \tn\ up to around 25~K with the same low-temperature value of $\gamma^{\rm mag}_c$. In general, at $T> T_{\rm N}$, the temperature variation of $\gamma^{\rm eff}_i$ in Fig.~\ref{fig:totalgruen} implies the persistent dominance of magnetic degrees of freedom and further corroborates the presence of significant short-range magnetic order. The values of $\gamma_i$ descend monotonically upon heating up to about 60~K which we associate with decreasing short-range correlations. The effective Gr\"{u}neisen parameters hence indicate short-range magnetic order up to $\gtrsim 4\times T_{\rm N}$ which agrees with the extended regime of magnetic entropy changes seen in the magnetic specific heat (cf. Fig.~\ref{cpfisher}) and to the magnetic susceptibility~\cite{Hergett2025}. This reflects the quasi-2D nature of magnetism in \lfs. We also note that magnetic Gr\"{u}neisen scaling by the low-temperature parameters clearly fails for the $a$ and $b$ axes at $T>T_{\rm N}$ (Fig.~\ref{fig:gruen2}b and c). This rules out that short-range order above \tn\ is driven by the same energy scale driving long-range order and suggests the presence of competing degrees of freedom. This is particularly evident in the $b$ axis data: for $T\leq T_{\rm N}$, the data show $\partial T_{\rm N}/\partial p_{\rm b}>0$ which is unambiguously demonstrated by the sign of the anomaly at \tn. In contrast, there is a regime of \textit{negative} uniaxial pressure dependence $\partial T/\partial p_{\rm a}<0$ for $T> T_{\rm N}$.

Overall, the short-range ordered regime is highly anisotropic in its coupling to strain. The approximate persistence of the $c$ axis scaling above \tn\ suggests that $c$ axis strain remains primarily sensitive to the dominant magnetic energy scale. The breakdown of Gr\"{u}neisen scaling for the  $a$ and $b$ axes above \tn\ appears close to the dimensional-crossover regime inferred from the critical behavior~\cite{Hergett2026}. This crossover occurs around T~$\simeq$ 22~K, where the system evolves from three-dimensional to two-dimensional Ising-like behavior at higher temperatures. From the observed failure of low-temperature Gr\"{u}neisen scaling above \tn\ for the $a$ and $b$ axis one might conclude the presence of a competing interaction which is particularly susceptible to strain along either of these two directions.

The Gr\"{u}neisen parameters and pressure dependencies resulting from the scaling at $T<$~\tn, illustrated in Fig.~\ref{fig:gruen2}, are listed in Table~\ref{tab:gruen}. The table also includes the pressure dependencies of \tn\ obtained by means of the Souza-Pippard analysis which exploits the actual  anomalies at \tn\ (see Fig.~\ref{lambda_alpha_cp}). Specifically, the Gr\"{u}neisen analysis yields the pressure dependence of the dominant magnetic energy scale in the temperature-regime under study (Eq.~\eqref{eq:single}), whereas the Souza-Pippard analysis yields the pressure dependence of the phase boundary, $\partial T_{\rm N}/\partial p_i$.

\begin{table}[htb]
\centering
\caption{
Magnetic Gr\"{u}neisen parameters $\gamma_i^{\rm mag}$ obtained from the low-temperature ($T<T_{\rm N}$) scaling shown in Fig.~\ref{fig:gruen2}, together with the corresponding pressure dependencies of the dominant energy scale $\left(\partial \ln \varepsilon/\partial p_i\right)_{\rm Gr}$. The last column shows the pressure dependencies of \tn\ obtained from the Souza-Pippard analysis.}

\begin{tabular}{c c c c}
\hline\hline
 &
$\gamma_i^{\rm mag}$ &
$\left(\partial \ln \varepsilon/\partial p_i\right)_{\rm Gr}$ &
$\left(\partial \ln T_{\rm N}/\partial p_i\right)_{\rm SP}$ \\
 &
$(10^{-6}\,\mathrm{mol/J})$ &
$(\%/\mathrm{GPa})$ &
$(\%/\mathrm{GPa})$ \\
\hline
$a$ &
-0.2(1) &
-1.0(5) &
-4.2(1) \\

$b$ &
0.2(1) &
1.0(5) &
6.3(2) \\

$c$ &
1.8(1) &
9.1(5) &
15.2(7) \\

vol &
3$\times$0.6(1) &
3$\times$3.0(5) &
14.5(5) \\
\hline\hline
\end{tabular}
\label{tab:gruen}
\end{table}

\subsubsection{Separating phonon and orbital degrees of freedom}

\begin{figure}[hb]
    \centering
    \includegraphics[width=\columnwidth]{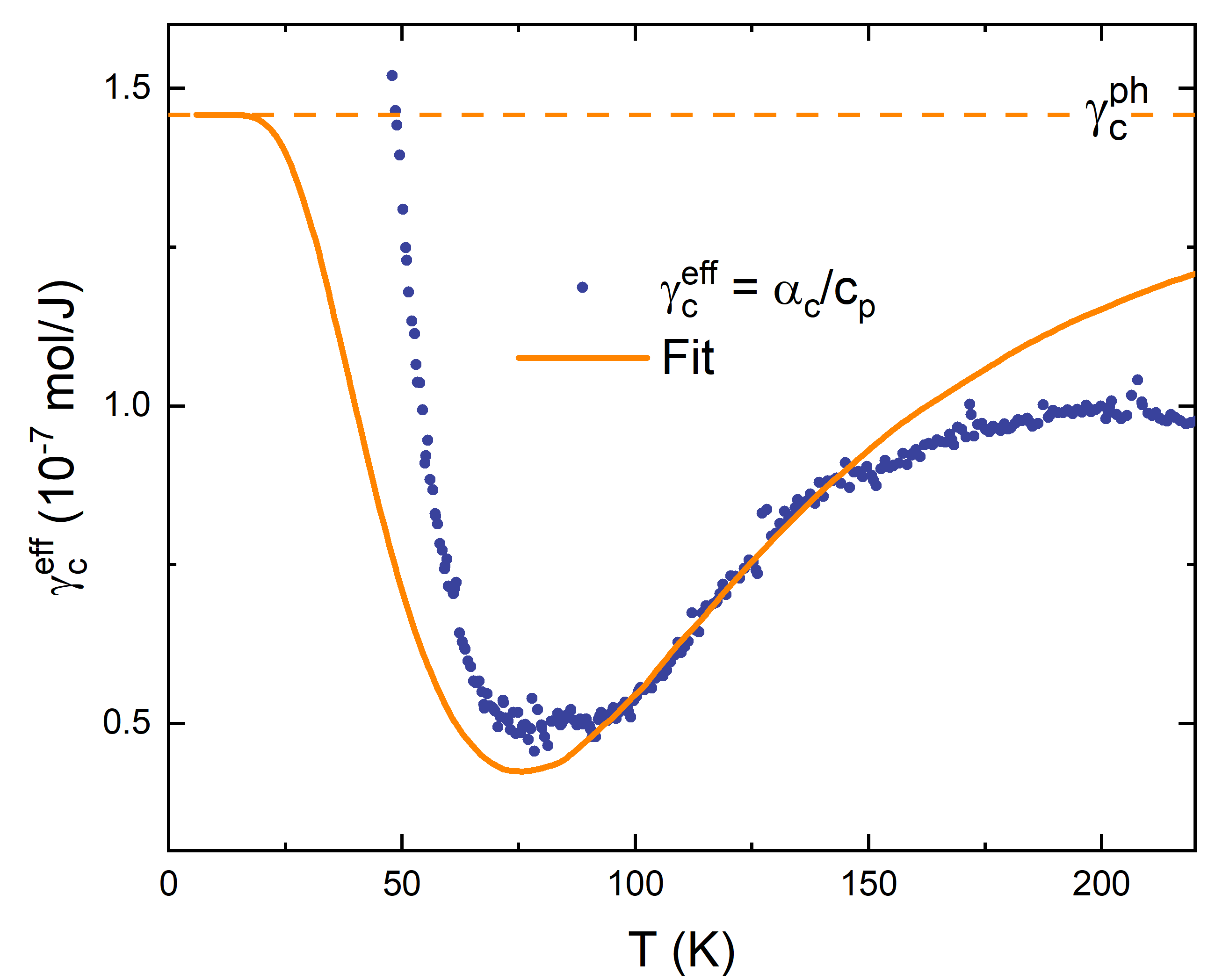}
    \caption{Effective $c$ axis-Gr\"{u}neisen parameter $\geffc=\alpha_{\rm c}/c_{\rm p}$ (blue symbols). The orange solid line shows a fit by means of Eq.~\ref{gammafit} (see the text for details). The dashed horizontal line indicates the phonon contribution $\gcph$ obtained from the fit.}
    \label{schottky_gamma}
\end{figure}


Following the same scheme applied to discriminate the spin and lattice degrees of freedom at low temperatures, the comparison of the thermal expansion coefficients and the specific heat enables us to further identify the orbital contributions to the thermodynamic response and to quantify the pressure dependence of the relevant energy scale, i.e., the energy gap $\Delta$ between the two lowest orbital states. Fig.~\ref{schottky_gamma} shows the temperature dependence of the measured total $c$ axis-Gr\"{u}neisen parameter $\geffc = \alpha_{\rm c}/c_{\rm p}$ calculated from the bare experimental data. For the following analysis, we use Eq.~\ref{eq:gamma_eff} and start considering the spin, lattice, and orbital degrees of freedom, i.e.,
\[
\gamma_c^{\mathrm{eff}}
=
\frac{
c_p^{\mathrm{mag}}\gamma_c^{\mathrm{mag}}
+
c_p^{\mathrm{ph}}\gamma_c^{\mathrm{ph}}
+
c_p^{o}\gamma_c^{o}
}{
c_p^{\mathrm{mag}} + c_p^{\mathrm{ph}} + c_p^{o}
},
\]
where $\gci$ and $c^{i}_{\rm p}$ are the magnetic, phononic, and orbital Gr\"{u}neisen parameters and specific heat contributions of the relevant degrees of freedom. As discussed in detail in Sec.~\ref{lro_sro}, magnetic correlations fade out above $\sim 60$~K so that for higher temperatures the effective Gr\"{u}neisen parameter is dominated by the lattice and orbital contributions. We may therefore write
\[
\gamma_c^{\mathrm{eff}}
=
\frac{
c_p^{\mathrm{ph}}\gamma_c^{\mathrm{ph}}
+
c_p^{o}\gamma_c^{o}
}{
c_p^{\mathrm{ph}} + c_p^{o}
}
=
\gamma_c^{\mathrm{ph}}
+
\left(\gamma_c^{o}-\gamma_c^{\mathrm{ph}}\right)
\frac{c_p^{o}}{c_p^{\mathrm{ph}}+c_p^{o}} .
\]

The phonon contribution to the specific heat, \cpph, obtained by temperature rescaling of \cpzn\ with $\rho_{\mathrm{D}} \approx 1.029$, is shown in Fig.~\ref{app:cpph_zn} in the Appendix. Since \cpo(T)$\ll$ \cpph(T)~\footnote{E.g., at 100~K, \cpo~$\approx 3.2~\mathrm{J\,mol^{-1}\,K^{-1}}$ and \cpph $\approx 40.9~\mathrm{J\,mol^{-1}\,K^{-1}} $ (see Fig.~\ref{cpfisher} and Fig.~\ref{app:cpph_zn} in the Appendix).}, $\geffc(T)$ can be expanded to leading order in the small parameter \cpo(T)/\cpph(T), yielding

\[
\gamma_c^{\mathrm{eff}}(T)
\approx
\gamma_c^{\mathrm{ph}}
+
\left(\gamma_c^{o}-\gamma_c^{\mathrm{ph}}\right)
\frac{c_p^{o}(T)}{c_p^{\mathrm{ph}}(T)} .
\]
Using Eq.~\ref{eq:schottky} for the orbital specific-heat contribution, \cpo(T), we obtain the following phenomenological description of the temperature dependence of $\gamma^{\rm eff}_c$:

\begin{equation}
\gamma_c^{\mathrm{eff}}(T)
\approx
\gamma_c^{\mathrm{ph}}
+
\left(\gamma_c^{o}-\gamma_c^{\mathrm{ph}}\right)
\frac{
R\left(\frac{\Delta}{T}\right)^2 e^{-\Delta/T}
}{
c_p^{\mathrm{ph}}(T)\left(1+e^{-\Delta/T}\right)^2
}.
\label{gammafit}
\end{equation}

We fit Eq.~\ref{gammafit} to the data above 90~K, i.e., the temperature regime where we expect vanishing magnetic contributions to the effective Gr\"{u}neisen parameter, using $\Delta$, $\gcph$, and $\gco$ as free parameters. \rev{The upper fit limit of 150~K is set by the assumption of a temperature-independent phononic Gr\"{u}neisen parameter, which fails once higher-energy optical modes become populated (see the Appendix for the sensitivity of the fit to the choice of the fitting regime).}





\rev{From the fit, we obtain $\Delta = 290(20)$~K, $\gcph = 14.6(17) \times 10^{-8}\,\mathrm{mol/J}$ and $\gco = -9.2(24) \times 10^{-7}\,\mathrm{mol/J}$, where the uncertainties are dominated by the systematic effect of varying the fit boundaries within the admissible window and exceed the statistical fitting errors by a factor of 3-5 (see the Appendix).} The obtained gap is reasonably close to the gap value derived from the Schottky analysis of the specific heat using Debye scaling. \rev{Imposing $\Delta = 330$~K on the fit describes the data at similar quality, so that both determinations are compatible within the stated uncertainties.} As described, higher-energy Einstein modes are expected from modeling the phonon specific heat. Once these modes acquire appreciable thermal weight, the phonon Gr\"{u}neisen parameter $\gcph$ may become temperature dependent. This effect is not included in Eq.~\ref{gammafit} and \rev{ detrmines the upper bound of the fitting window.}

Using the molar volume at 100~K, $V_m = VN_A/Z \approx 50.55~\mathrm{cm^3\,mol^{-1}}$, obtained from the unit-cell volume $V=335.732$~\AA$^3$ and $Z=4$ formula units per unit cell~\cite{Hergett2019b}, the uniaxial pressure dependence of the orbital gap is
\[
\left.\frac{\partial \ln \Delta}{\partial p_c}\right|_T
=
V_m \gco
=
-4.6(12)~\%/\mathrm{GPa}
\]
For $\Delta=290(20)$~K, this yields the pressure parameter $\partial \Delta/\partial p_{\rm c} = \Delta \times \partial \ln\Delta/\partial p_{\rm c} \simeq -13(5)~\mathrm{K/GPa}$.

\begin{figure}[htb]
    \centering
    \includegraphics[width=\columnwidth]{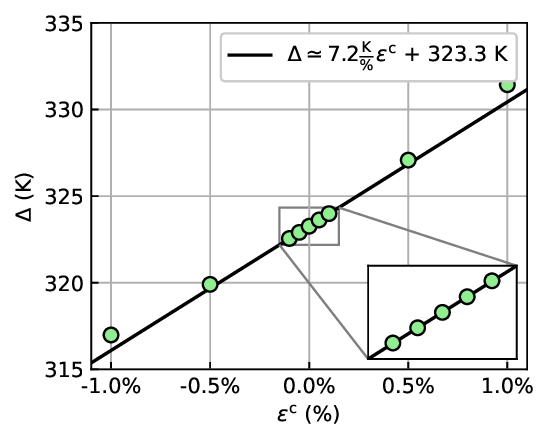}
    \caption{Uniaxial strain dependence of the energy gap $\Delta$ between the two lowest orbital levels. Strain, $\varepsilon^c$, is applied along the crystallographic $c$ axis, while the lattice parameters $a$ and $b$ are scaled equally to preserve the unit-cell volume. The black line represents a linear fit, indicating an approximately linear dependence $\Delta(\varepsilon^c)$.}
    \label{delta-strain}
\end{figure}


The experimental value of $\partial \Delta/\partial p_{\rm c}$ can be compared
with the theoretical value obtained from DFT. Within the local-density approximation,
we find an approximately linear dependence of the orbital gap $\Delta$ on strain applied
along the crystallographic $c$ axis. At the experimental equilibrium structure,
\[
\frac{\partial \Delta}{\partial \varepsilon^c}
\simeq 7~\mathrm{K}/\% ,
\]
where $\varepsilon^c$ denotes the $c$-axis strain expressed in percent.
To convert this strain derivative into a pressure derivative, we use the compressibility
reported for the $Pmnb$ phase of \lfs\ in Ref.~\cite{Vajeeston2017},
$\kappa \simeq 0.014~\mathrm{GPa}^{-1}$. Assuming
$\varepsilon^c[\%] \simeq -100\,\kappa p_{\rm c}$, we obtain
\[
\frac{\partial \Delta}{\partial p_{\rm c}}
\simeq -10~\mathrm{K/GPa}.
\]
Thus, the DFT-based estimate agrees with the experimental result within about $25~\%$.
Given the approximations involved, this level of agreement is quite satisfactory. In particular, crystal-field splittings obtained from DFT are evaluated in the
self-consistent potential of the ground state. For charge-neutral local excitations,
this can introduce sizable errors, because part of the local Coulomb response to the
orbital excitation is already implicitly included in the ground-state potential.

\section*{Summary}

In summary, our detailed studies of the thermal expansion and the specific heat of Li$_2$FeSiO$_4$ single crystals enable us to investigate magneto-elastic coupling and in particular allow us to separate the spin, lattice, and orbital degrees of freedom in the magnetically quasi-2D material. Our analysis yields strongly anisotropic uniaxial strain dependencies which are dominated by one (i.e., the $c$ axis) in-plane uniaxial pressure dependence $(\partial \ln{ T_{\rm N}} )/\partial p_{\rm c} \simeq 15.2(7)$~\%/GPa. The (negative) $a$ axis and (positive) out-of-plane ($b$ axis) pressure dependencies are much smaller. The non-monotonic temperature dependence of the in-plane distortion, a sign change of the Gr\"{u}neisen parameter $\gamma_{\rm b}$, and the failure of Gr\"{u}neisen scaling imply the presence of competing interactions in the magnetically short-range ordered phase which extends to $\gtrsim 4\times T_{\rm N}$. The onset of this regime of competing interactions corresponds to the crossover from approximately 2D Ising-like to 3D behavior. Both the specific heat data and the Gr\"{u}neisen ratio clearly evidence the relevance of orbital excitations. The lowest orbital excitation gap, as well as its uniaxial pressure dependence, is extracted from the experimental data. Both results agree very well with numerical predictions by means of DFT. Our data illustrate how long-range magnetic order evolves in a 2D Ising-like system in the presence of orbital degrees of freedom.

\begin{acknowledgments}

We acknowledge the support by Deutsche  Forschungsgemeinschaft (DFG) under Germany’s Excellence Strategy EXC2181/1-390900948 (The Heidelberg STRUCTURES Excellence Cluster). We are indebted to Siegmar Roth and Andre Beck at the Institute for Quantum Materials and Technologies, Karlsruhe Institute of Technology, for their technical assistance and support during the XRD measurements.
\end{acknowledgments}

\bibliography{Grueneisen_biblio}

\hfill
\newpage

\renewcommand{\floatpagefraction}{1.0}
\renewcommand{\topfraction}{1.0}
\renewcommand{\bottomfraction}{1.0}
\renewcommand{\textfraction}{0.0}

\section*{Appendix}
\label{sec:appendix}

\renewcommand{\thefigure}{A\arabic{figure}}
\setcounter{figure}{0}

\renewcommand{\thesection}{A\arabic{section}}
\setcounter{section}{0}



\begin{figure}[b]
\begin{center}
\includegraphics[width=0.95\columnwidth,clip]{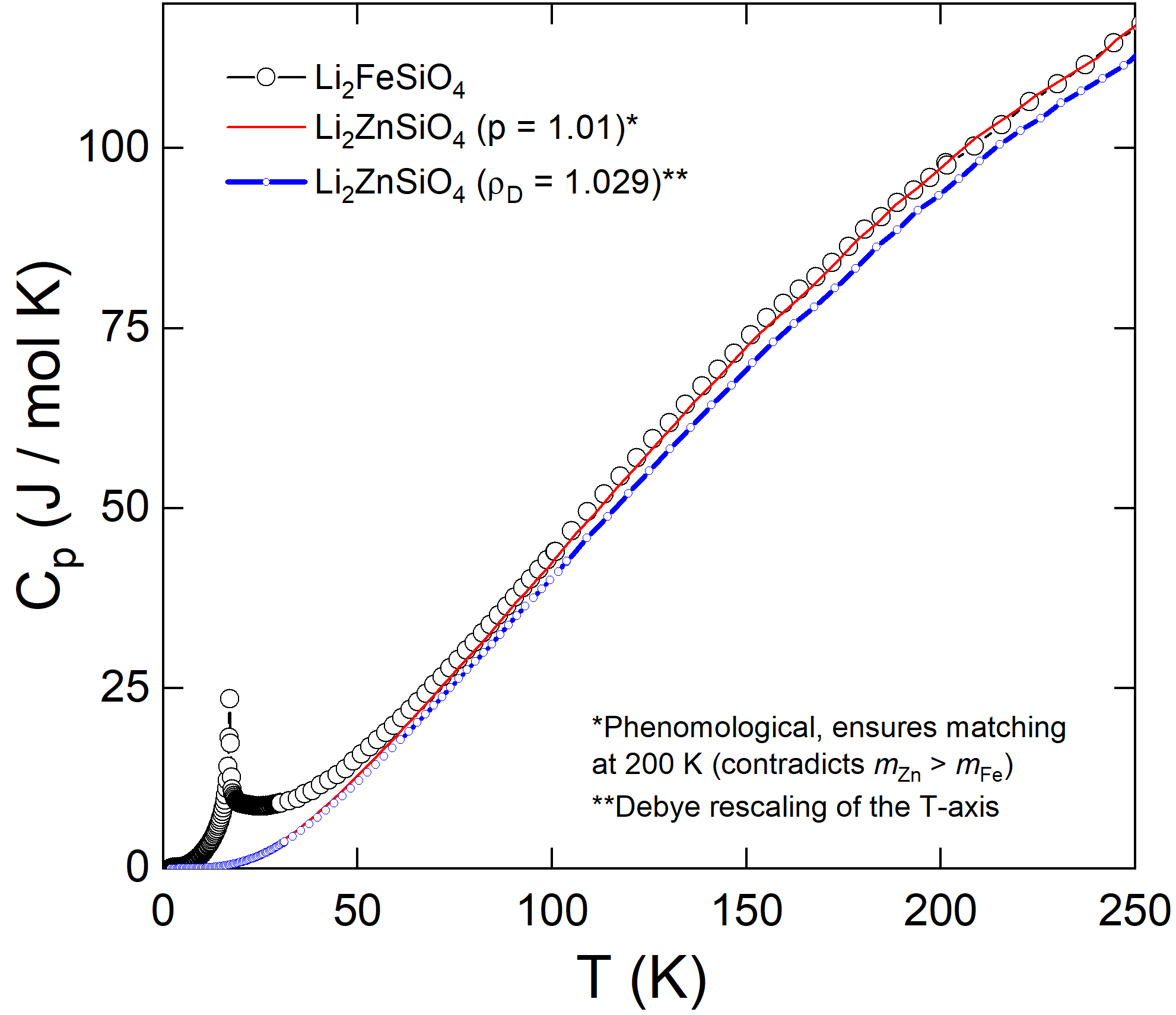}
\caption{\rev{Illustration of obtaining the phonon specific heat \cpph\ from the experimental specific-heat data on \lzs . The blue data (**) show the results of Debye rescaling the temperature axis using $\rho_{\mathrm{D}} = 1.029$.
The red line (*) illustrates a phenomenological approach previously used in Ref.~\cite{Hergett2025} which assumes that the specific heat at 200~K is purely phononic, i.e. \cpfe~=~\cpzn\ in this temperature regime. The latter approach results in an unreasonable scaling factor $p>1$ for \cpzn\ } and per construction disregards the high-temperature entropy changes associated with the lowest orbital excitation.}
\label{app:cpph_zn}
\end{center}
\end{figure}

\rev{The phonon background of the specific heat of \lfs\ is obtained from the measured specific heat of its non-magnetic analog \lzs . A quantitative approach to account for the different molar masses ($M$) of \lfs\ and \lzs\ is to renormalize the data using the Debye-temperature ratio $\rho_{\mathrm{D}} = \Theta_D(\mathrm{Li_2FeSiO_4}) / \Theta_D(\mathrm{Li_2ZnSiO_4})$ (see Fig.~\ref{app:cpph_zn} illustrating this Debye-renormalization). The phonon background is then evaluated by multiplying the temperature values by this ratio in the \cpzn\ versus $T$ curve. Within the commonly adopted approximation (see, e.g., Ref.~\cite{Wang2012}), the Debye-temperature ratio is estimated as $\rho_{\mathrm{D}} \approx \sqrt{M_{\mathrm{Li_2ZnSiO_4}}} / \sqrt{M_{\mathrm{Li_2FeSiO_4}}} = 1.029$.}

\begin{figure}[tb]
\begin{center}
 \includegraphics[width=0.9\columnwidth,clip]{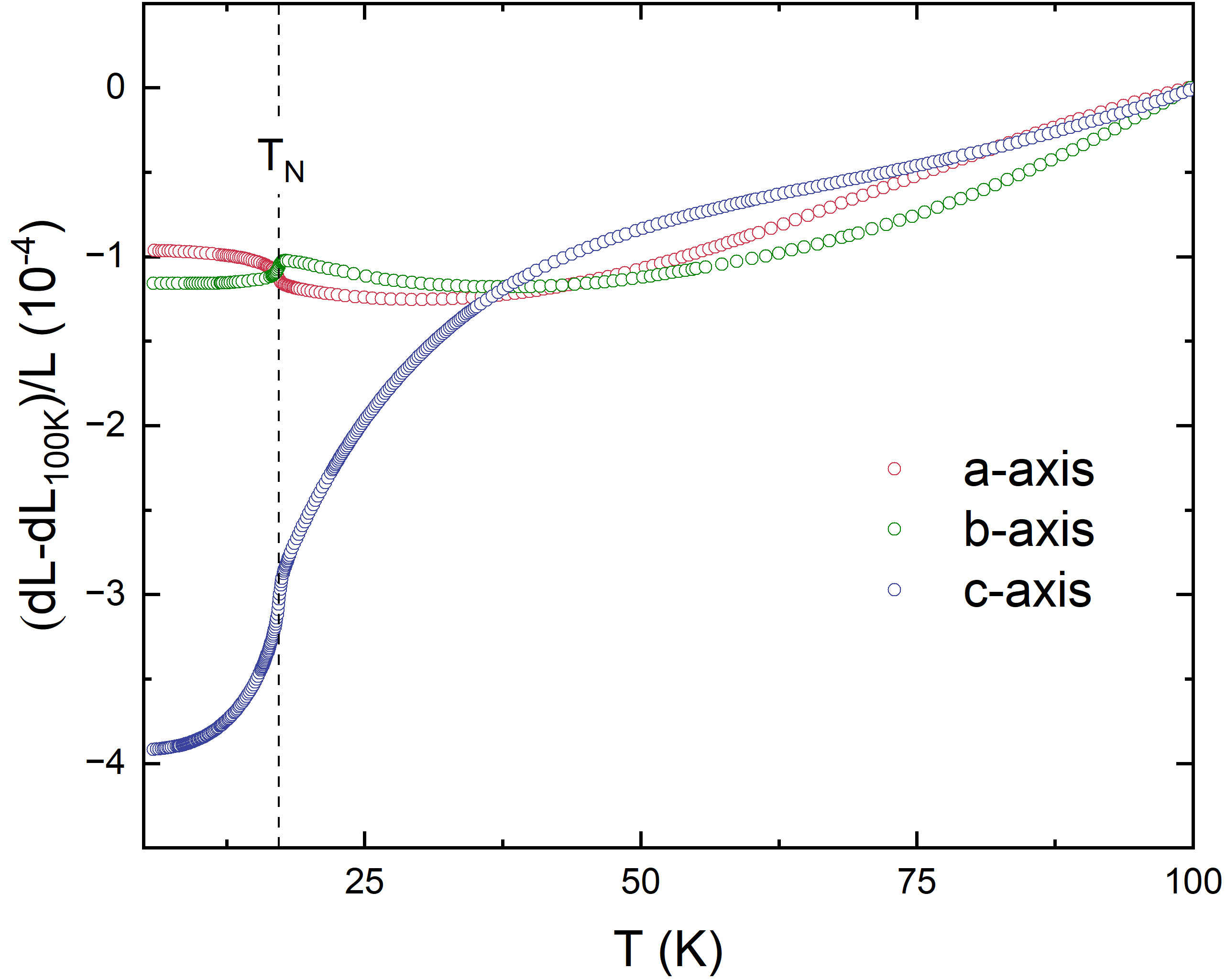}
 \caption{Temperature dependence of the relative length changes along the crystallographic $a$, $b$, and $c$ axes, at $T\leq 100$~K, referenced to their respective values at 100~K.}
\label{fig:SM_dll_shift}
\end{center}
\end{figure}

\rev{Note that this procedure differs from the purely phenomenological method reported in Ref.~\cite{Hergett2026} where \cpphon\ was estimated by matching the specific heat data of \lfs\ and \lzs\ above $\sim 150$~K, i.e., at the temperatures where Fisher's specific heat implies the absence of short-range magnetic order. The previously obtained scaling factor ($p=1.01$) for \cpzn\ (i.e., \cpphon = $p\,$\cpzn), however, contradicts the expected scaling factor $p<1$ which is inferred from the higher molar mass of Zn (65.38~g/mol) compared to Fe (55.85~g/mol). Note that $\rho_{\rm D}>1$ from Debye scaling agrees with $p<1$. The phenomenological approach yields a reasonable estimate of the bare spin entropy changes. However, per construction it disregards the high-temperature entropy changes associated with the lowest orbital excitation. The relative length changes along the crystallographic $a$, $b$, and $c$ axes, referenced to their respective values at 100~K, are shown in Fig.\ref{fig:SM_dll_shift} for the same temperature regime.
}




\begin{figure}[t]
\begin{center}
 \includegraphics[width=0.99\columnwidth,clip]{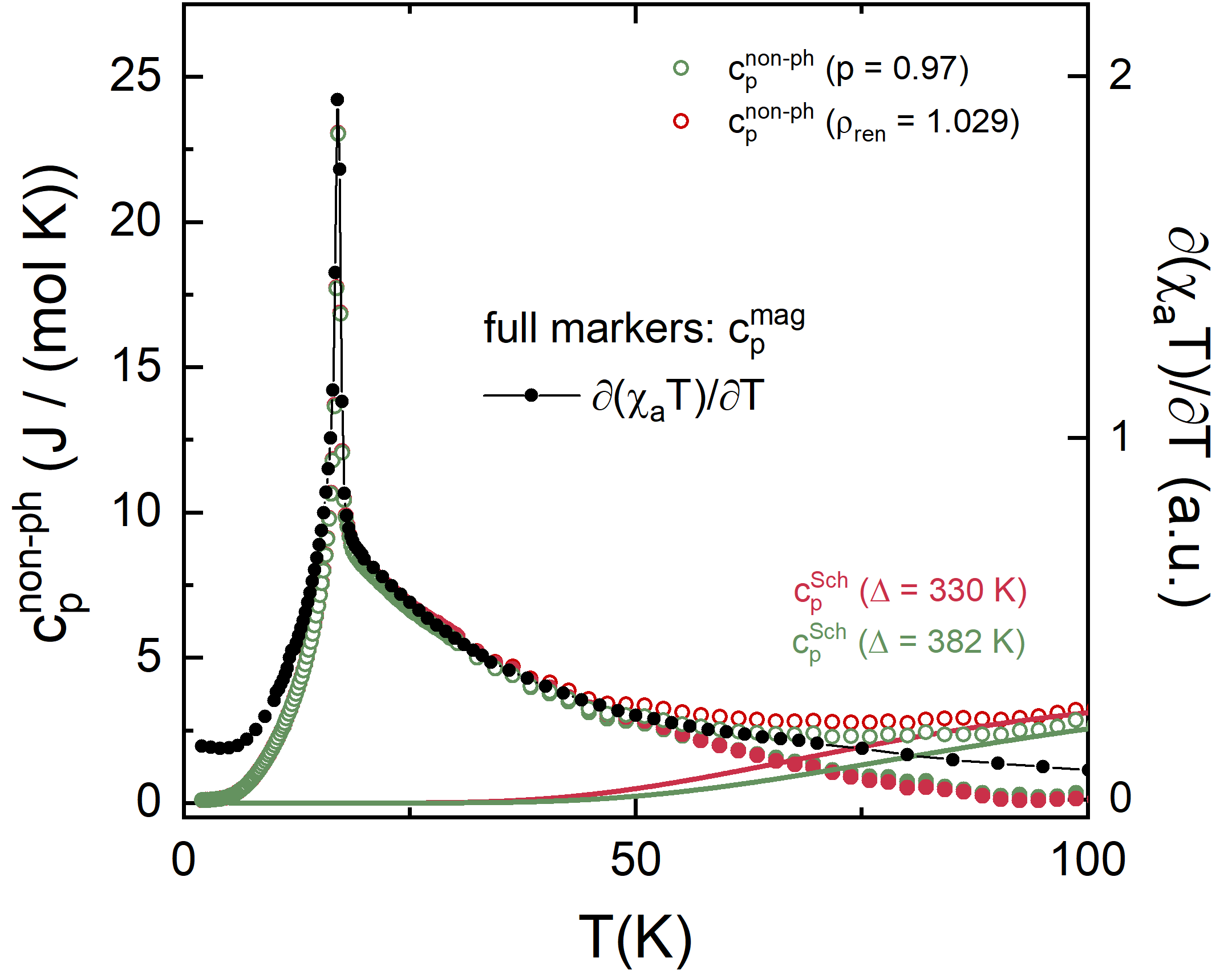}
 \caption{\rev{Non-phonon specific heat (open data points) and estimate of the Schottky (\cpsch ) and magnetic (\cpmag ) specific heat taken from Fig.~\ref{cpfisher} for $T<100$~K. Black data show that Fisher's specific heat from Fig.~3b excellently agrees with \cpmag .}}
\label{fig:appendix-fisher}
\end{center}
\end{figure}



\begin{figure}[h]
\begin{center}
 \includegraphics[width=0.9\columnwidth,clip]{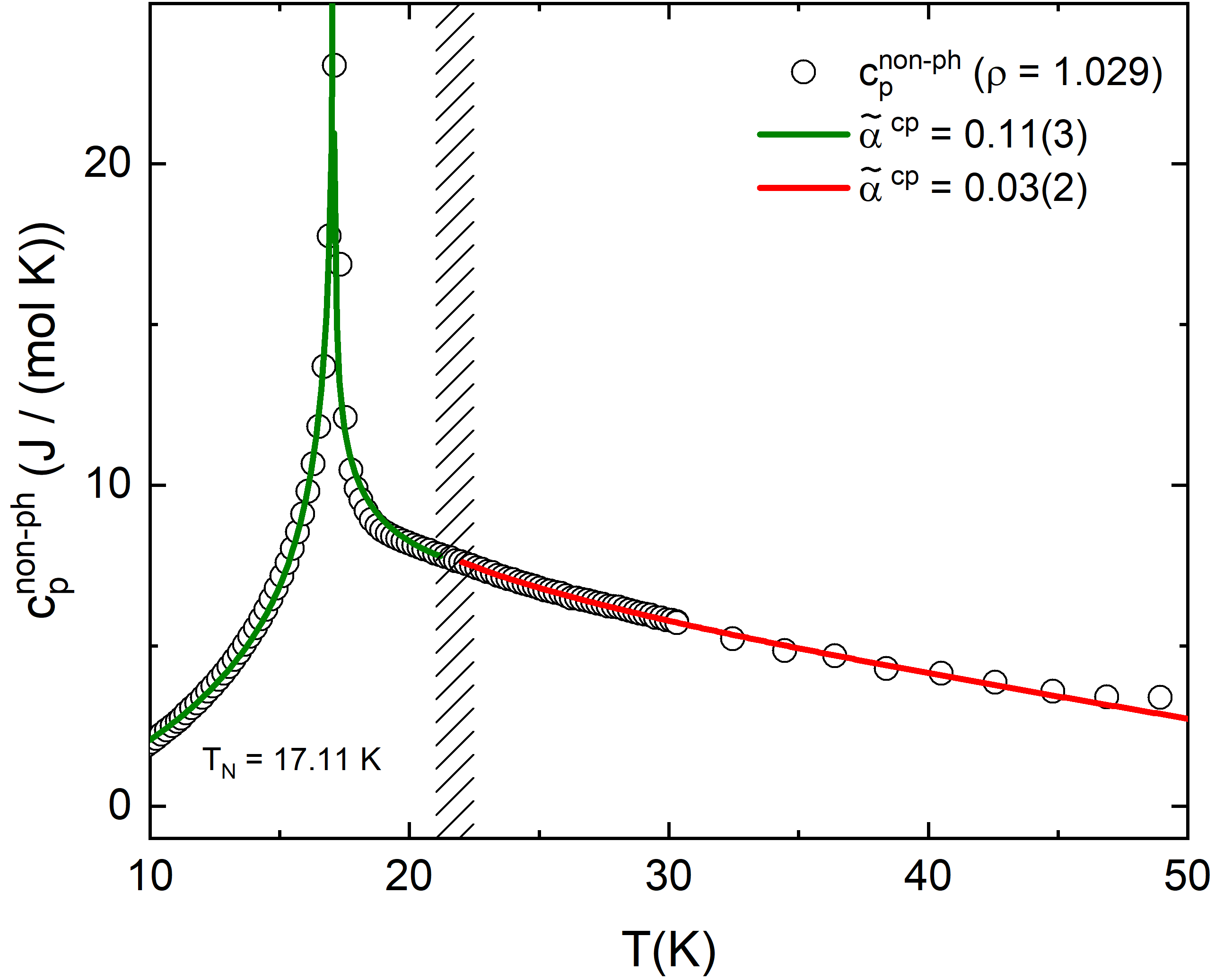}
 \caption{Non-phononic heat capacity, \cpnon\ (open circles), obtained by temperature rescaling and subtraction of \cpzn\ with $\rho_{\mathrm{D}} \approx 1.029$ (see the main text). The solid green and red lines are fits to the data in different temperature regimes, yielding the critical exponents $\tilde{\alpha}^{\rm cp}=0.11(3)$ (around \tn ) and $\tilde{\alpha}^{\rm cp}=0.03(2)$, respectively. The hatched area marks the 2D-3D crossover region (see Ref.~\cite{Hergett2026}).}
\label{app:critical}
\end{center}
\end{figure}

The quantitative approach (Debye scaling) to obtain the non-phononic heat capacity yields slightly different results than the phenomenological approach in Ref.~\cite{Hergett2026}. We hence have revisited the critical analysis of the specific-heat-anomaly at \tn . The results do not significantly differ from the previous reported values and are presented in Fig.~\ref{app:critical}.

\begin{figure}[h]
    \centering
    \includegraphics[width=0.85\columnwidth, clip]{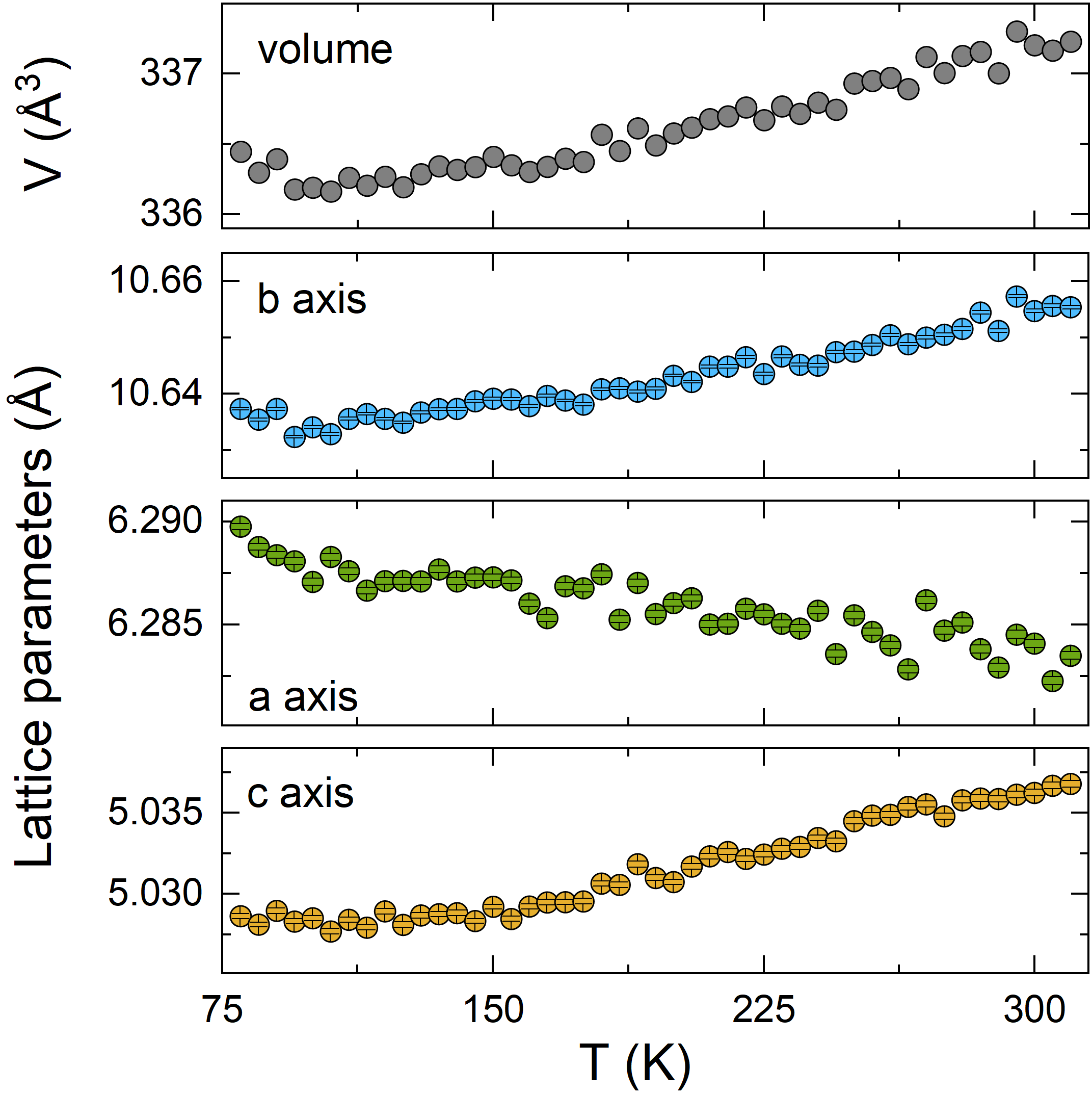}
    \caption{Lattice parameters (in $Pmnb$ setting) and unit cell volume in the temperature range 80~K~$\leq T\leq 310$~K derived from single-crystal x-ray diffraction measurements. }
    \label{fig:xrd1}
\end{figure}

\begin{figure}[t]
\begin{center}
 \includegraphics[width=0.9\columnwidth,clip]{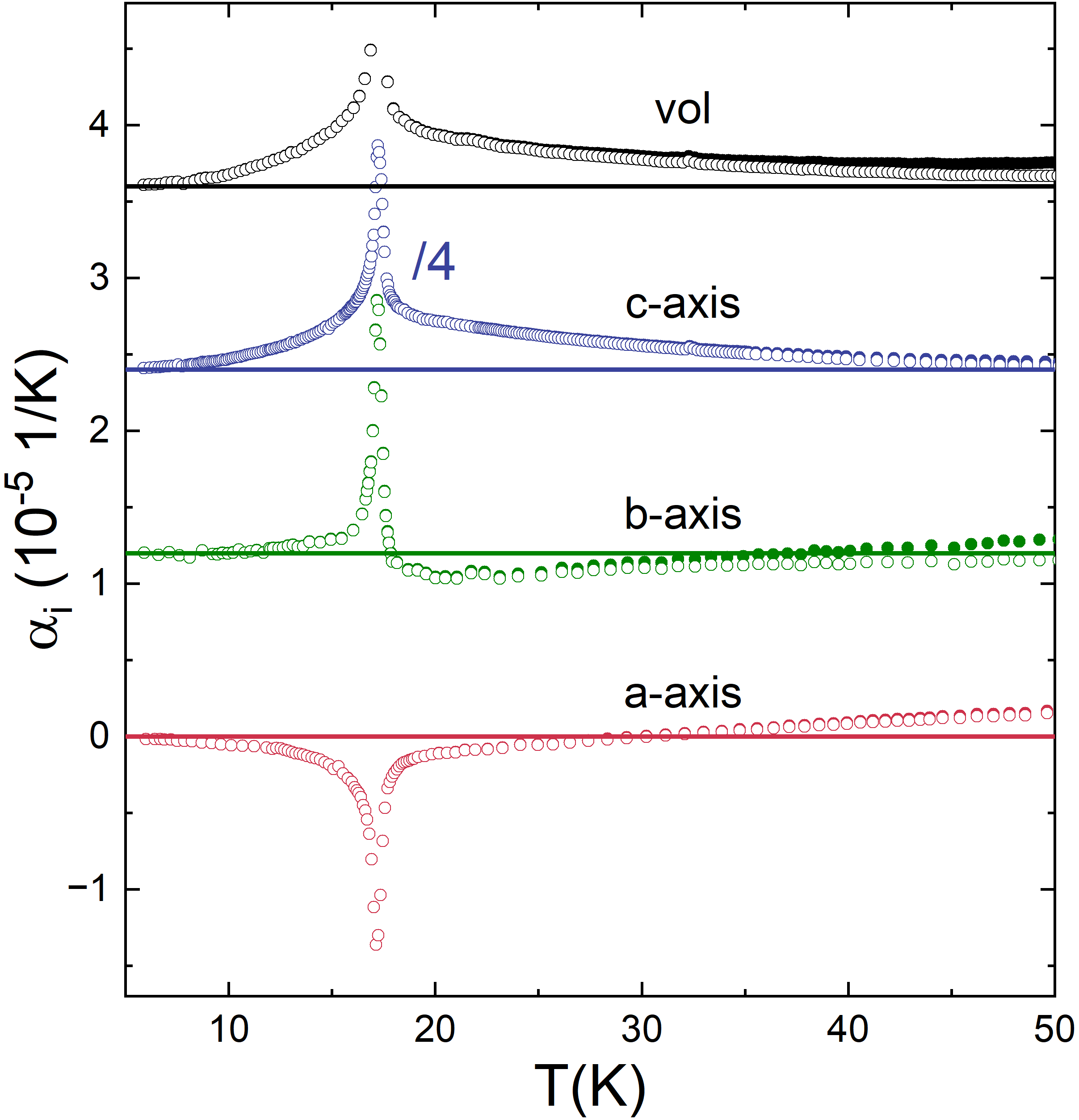}
 \caption{The negligible effect of background correction on the uniaxial and volume thermal expansion coefficients at low temperatures: solid symbols represent the measured data, open symbols show the background-corrected data using the scaling factor $\rho_{\mathrm{D}} \approx 1.029$ as described in the main manuscript text. The data have been shifted vertically by $1.2\times 10^{-5}$/K for clarity; horizontal lines show $\alpha_{\rm i}=0$ for each data set; \alc\ has been divided by 4.
 }
\label{fig:TE-background}
\end{center}
\end{figure}

\begin{figure}[t]
\begin{center}
 \includegraphics[width=0.9\columnwidth,clip]{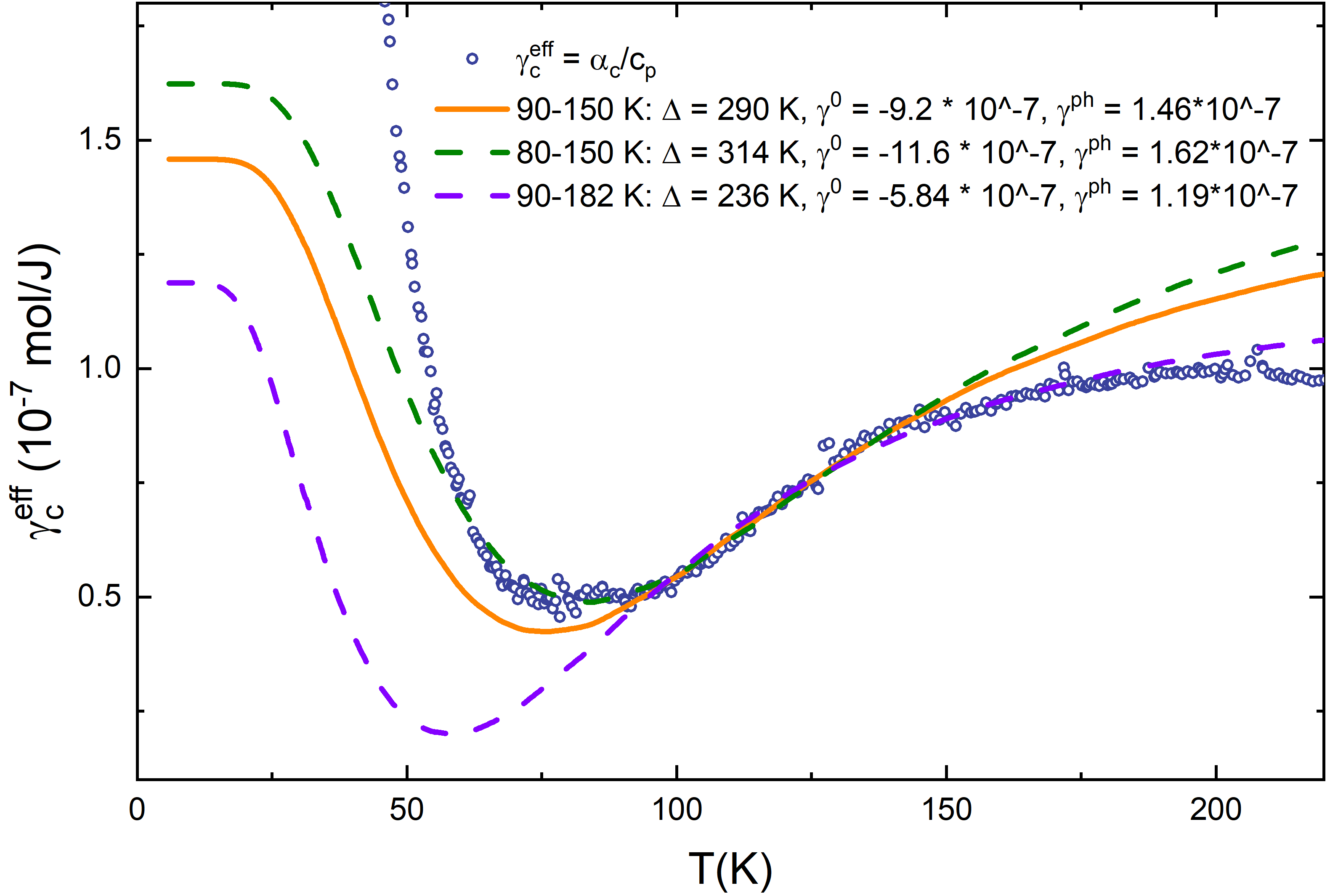}
 \caption{Effective $c$ axis Gr\"{u}neisen parameter $\gamma_c^{\mathrm{eff}}=\alpha_c/c_{\rm p}$ as a function of temperature (open blue circles).  The lines show fits to Eq.~\eqref{gammafit} performed over different temperature intervals. 
  The solid orange curve corresponds to the fit range $90$-$150\,\mathrm{K}$ and yields $\Delta=290\,\mathrm{K}$, $\gamma_c^0=-9.2\times10^{-7}\,\mathrm{mol/J}$, and $\gamma_c^{\mathrm{ph}}=14.6\times10^{-8}\,\mathrm{mol/J}$. \rev{Note that the 80-150 K and 90-182 K ranges violate the lower and upper validity condition, respectively (see text).}
}
\label{fig:Fits_g0_gph_Delta}
\end{center}
\end{figure}

Fig.~\ref{fig:Fits_g0_gph_Delta} illustrates the sensitivity of the extracted gap $\Delta$, of the orbital Gr\"{u}neisen parameter $\gamma_c^0$, and of the phononic Gr\"{u}neisen parameter $\gamma_c^{\mathrm{ph}}$ to the chosen fitting range. \rev{The interval 90-150~K used in the main text is bounded on both sides by physical conditions. Below $\sim$~90-100~K, magnetic contributions to $\gamma_c^{\mathrm{eff}}$ are finite, as evidenced by \cpnon~-~\cpsch~$>$~0 (Fig.~\ref{cpfisher}), while above $\sim$~150~K the assumption of a temperature-independent $\gcph$ fails once higher-energy optical modes become populated. A Debye-Einstein description of the specific heat yields $\Theta_{\mathrm{D}} = 343$~K and $\Theta_{\mathrm{E}} = 693$~K, so that appreciable population sets in above $\sim$~$\Theta_{\mathrm{E}}/5 \approx 140$~K, where systematic deviations are indeed observed (Fig.~\ref{schottky_gamma}). Both bounds are corroborated by the fits. Extending the fitting regme  below 90~K degrades the fit quality ($R^2 = 0.969$ for 80-150~K vs. 0.986 for 90-150~K). Extending the upper bound beyond 150~K shifts $\Delta$ systematically downwards (260~K and 236~K for upper bounds of 160 and 182~K) at nearly unchanged $R^2$, i.e., the increasing temperature dependence of $\gamma_c^{\mathrm{ph}}$ is absorbed into a smaller apparent gap. Free fits over the admissible fitting regimes 90-140~K and 90-150~K yield $\Delta$~=~307(5)~K and 290(6)~K at practically identical fit quality ($R^2 = 0.9863$ and 0.9861), the two values differing by three times the statistical error. The systematic uncertainties quoted in the main text cover the admissible fitting ranges.}

\end{document}